\RequirePackage{fix-cm}
\documentclass[twocolumn,epjc3]{svjour3}  
\smartqed  
\RequirePackage{graphicx}
\RequirePackage{orcidlink}
\RequirePackage{amsmath}
\RequirePackage{xspace}
\RequirePackage{comment}
\RequirePackage{upgreek}
\newcommand{\mhz}{\ensuremath{~\mathrm{MHz}}\xspace}
\newcommand{\cm}{\ensuremath{~\mathrm{cm}}\xspace}
\newcommand{\gevc}{\ensuremath{~\mathrm{GeV/}c}\xspace}
\newcommand{\gev}{\ensuremath{~\mathrm{GeV}}\xspace}

\newcommand{\mm}{\ensuremath{~\mathrm{mm}}\xspace}
\newcommand{\mum}{\ensuremath{~\mathrm{\mu m}}\xspace}
\newcommand{\pt}{\ensuremath{~\mathrm{p_{\text{T}}}}\xspace}

\newcommand{\second}{\ensuremath{~\mathrm{s}}\xspace}
\newcommand{\lhcb}{\text{\mbox{LHCb}}\xspace}
\newcommand{\lhc}{\text{\mbox{LHC}}\xspace}
\newcommand{\lum} {\ensuremath{\mathcal{L}}\xspace}

\newcommand{\hltone}{\text{\mbox{HLT1}}\xspace}
\newcommand{\hlttwo}{\text{\mbox{HLT2}}\xspace}
\newcommand{\velo}{\text{\mbox{VELO}}\xspace}

\newcommand{\bquark}{\textit{b}\xspace}
\newcommand{\cquark}{\textit{c}\xspace}
\newcommand{\proton}{\textit{p}\xspace}

\newcommand{\zpoca}{\ensuremath{z_\text{poca}}\xspace}
\newcommand{\xtrk}[1][]{\ensuremath{x_\text{trk #1}}\xspace}
\newcommand{\ytrk}[1][]{\ensuremath{y_\text{trk $#1$}}\xspace}
\newcommand{\ztrk}[1][]{\ensuremath{z_\text{trk $#1$}}\xspace}
\newcommand{\txtrk}{\ensuremath{t_{x,\text{trk}}}\xspace}
\newcommand{\tytrk}{\ensuremath{t_{y,\text{trk}}}\xspace}
\newcommand{\xtrki}[1][]{\ensuremath{x_\text{trk $i$}}\xspace}
\newcommand{\ytrki}[1][]{\ensuremath{y_\text{trk $i$}}\xspace}
\newcommand{\ztrki}[1][]{\ensuremath{z_\text{trk $i$}}\xspace}
\newcommand{\txtrki}{\ensuremath{t_{x,\text{trk $i$}}}\xspace}
\newcommand{\tytrki}{\ensuremath{t_{y,\text{trk $i$}}}\xspace}
\newcommand{\xvtx}{\ensuremath{x_\text{vtx}}\xspace}
\newcommand{\yvtx}{\ensuremath{y_\text{vtx}}\xspace}
\newcommand{\zvtx}{\ensuremath{z_\text{vtx}}\xspace}
\newcommand{\ud}{\mathrm{d}}

\newcommand{\pvx}{PV$_{x}$\xspace}

\newcommand{\pvz}{PV$_{z}$\xspace}
\newcommand{\xbeam}{\ensuremath{x_\text{b}}\xspace}
\newcommand{\ybeam}{\ensuremath{y_\text{b}}\xspace}
\newcommand{\txbeam}{\ensuremath{t_{x,\text{b}}}\xspace}
\newcommand{\tybeam}{\ensuremath{t_{y,\text{b}}}\xspace}
\newcommand{\chisqmax}{\ensuremath{\chi^2_\text{max}}}

\journalname{Submitted to Eur. Phys. J. C}
\begin{document}\sloppy

\title{A parallel algorithm for fast reconstruction of primary vertices on heterogeneous architectures}


\author{Agnieszka Dziurda\thanksref{addr1,cor1}\orcidlink{0000-0003-4338-7156}
    \and
    Maciej Giza\thanksref{addr1}\orcidlink{0000-0002-0805-1561}
    \and 
    Vladimir V. Gligorov\thanksref{addr2,addr3}\orcidlink{0000-0002-8189-8267}
    \and 
    Wouter Hulsbergen\thanksref{addr4}\orcidlink{0000-0003-3018-5707}
    \and
    Bogdan Kutsenko\thanksref{addr5}\orcidlink{0000-0002-8366-1167}
    \and 
    Saverio Mariani\thanksref{addr3,cor3}\orcidlink{0000-0002-7298-3101}
    \and
    Niklas Nolte\thanksref{addr6}\orcidlink{0000-0003-2536-4209}
    \and 
    Florian Reiss\thanksref{addr7,cor2}\orcidlink{0000-0002-8395-7654}
    \and 
    Patrick Spradlin\thanksref{addr8}\orcidlink{0000-0002-5280-9464}
    \and 
    Dorothea vom Bruch\thanksref{addr5}\orcidlink{0000-0001-9905-8031}
    \and 
    Tomasz Wojton\thanksref{addr1}\orcidlink{0000-0003-3053-4305}
}

\thankstext{cor1}{email: agnieszka.dziurda@cern.ch (corresponding author)} 
\thankstext{cor2}{email: florian.reiss@cern.ch (corresponding author)} 
\thankstext{cor3}{email: saverio.mariani@cern.ch (corresponding author)} 


\institute{Henryk Niewodniczanski Institute of Nuclear Physics Polish Academy of Sciences, Krak\'ow, Poland \label{addr1}
           \and
           LPNHE, Sorbonne Universit\'e, Paris Diderot Sorbonne Paris Cit\'e, CNRS/IN2P3, Paris, France \label{addr2}
           \and
           European Organization for Nuclear Research (CERN), Geneva, Switzerland  \label{addr3}
           \and
           Nikhef National Institute for Subatomic Physics, Amsterdam, Netherlands \label{addr4}
           \and
           Aix Marseille Univ, CNRS/IN2P3, CPPM, Marseille, France
           \label{addr5}
           \and 
           Massachusetts Institute of Technology, Cambridge, MA, United States \label{addr6}
           \and
           Physikalisches Institut, Albert-Ludwigs-Universit{\"a}t Freiburg, Freiburg, Germany \label{addr7}
           \and 
           University of Glasgow, Glasgow, United Kingdom \label{addr8}
}


\maketitle

\begin{abstract}
The physics programme of the \lhcb experiment at the Large Hadron Collider requires an efficient and precise reconstruction of the particle collision vertices. The \lhcb Upgrade detector relies on a fully software-based trigger with an online reconstruction  rate of 30\mhz, necessitating fast vertex finding algorithms. This paper describes a new approach to vertex reconstruction developed for this purpose. The algorithm is based on cluster finding within a histogram of the particle trajectory projections along the beamline and on an adaptive vertex fit. Its implementations and optimisations on x86 and GPU architectures and its performance on simulated samples are also discussed.
\end{abstract}

\section{Introduction}
\label{sec:Introduction}

For Run~3 (2022-2026) of the Large Hadron Collider (\lhc), the \lhcb Upgrade~I detector~\cite{LHCb-TDR-012,LHCb-DP-2022-002} is designed to take data at an instantaneous luminosity of $\lum = 2 \times 10^{33} \cm^{-2} \second^{-1}$. This is five times larger than in previous data-taking periods and corresponds to an average number of five visible interactions per proton-proton ($pp$) bunch crossing (``event"), denoted as $\mu = 5$. The detector also includes an improved fixed-target system, called SMOG2~\cite{LHCb-TDR-020,LHCb-PUB-2018-015,SMOG2_paper}, consisting of a storage cell confining target gas in a 20\cm-long region upstream of the nominal \textit{pp} interaction point. By exploiting the interaction between the \lhc beam protons and injected gas ($p$-gas), \lhcb is the only experiment at the \lhc capable of simultaneously acquiring $pp$ and $p$-gas collisions. To cope with the increased event rate, \lhcb has implemented its real-time data processing (``trigger") in a heterogeneous farm of Graphics Processing Unit (GPU) and Central Processing Unit (CPU) processors. The task of this full software trigger~\cite{LHCb-TDR-016,LHCb-TDR-021} is to process data from the detector at a frequency of up to 30 MHz. The raw data rate is reduced from 4~TB/s to around 10~GB/s and then recorded to permanent storage. The trigger is divided into two stages: the first one (\hltone) runs on GPUs and reduces the data rate by a factor of around 30; the second one (\hlttwo) runs on x86 CPU processors.
Each algorithm in the \lhcb event reconstruction software has been updated to achieve the desired event throughput and physics performance. This work, ongoing since 2015, has required \lhcb to overhaul its previously sequential reconstruction code, in order to exploit modern parallel computing architectures.

\begin{figure*}
\centering
\includegraphics[width=0.49\linewidth]{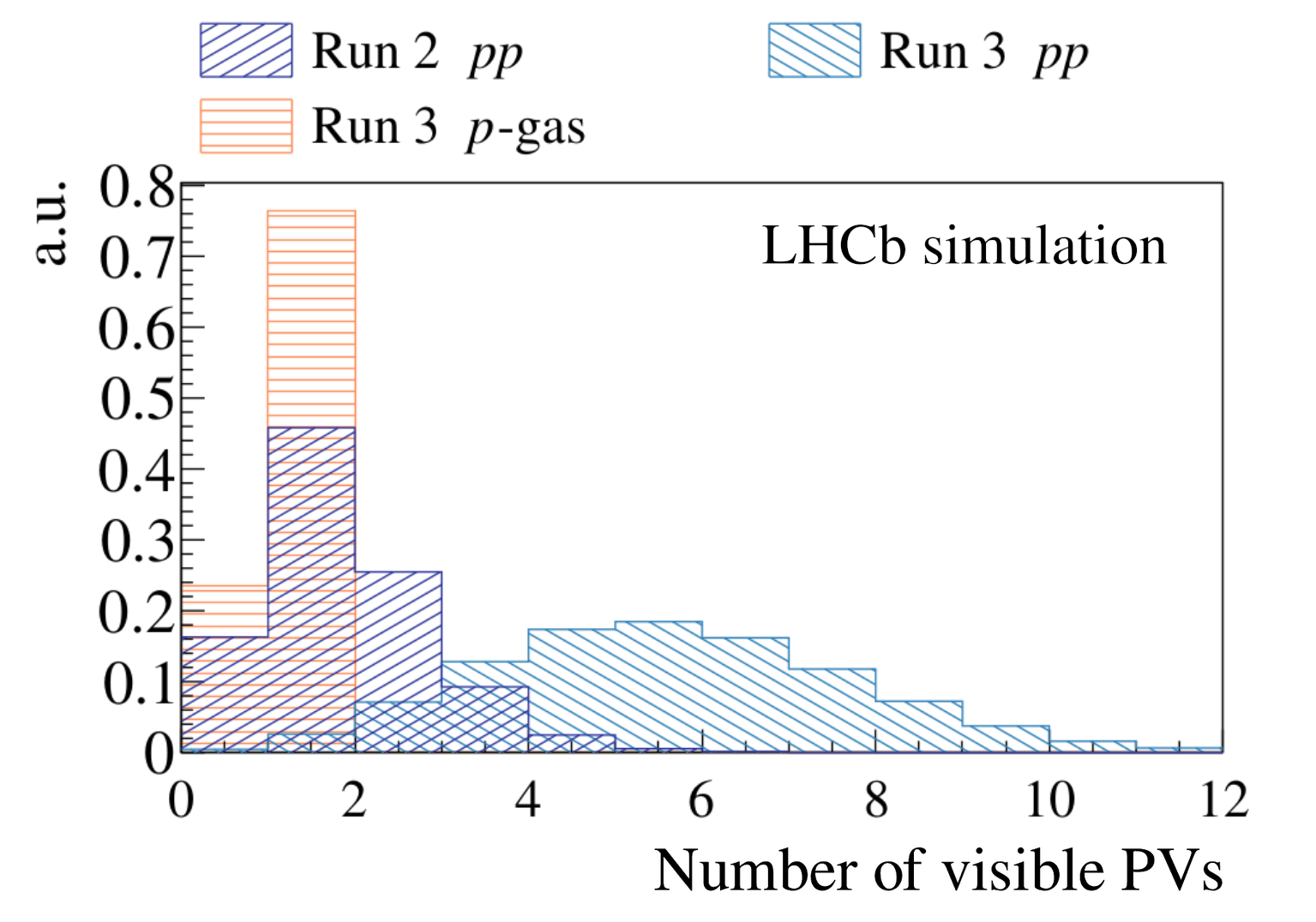}
\includegraphics[width=0.49\linewidth]{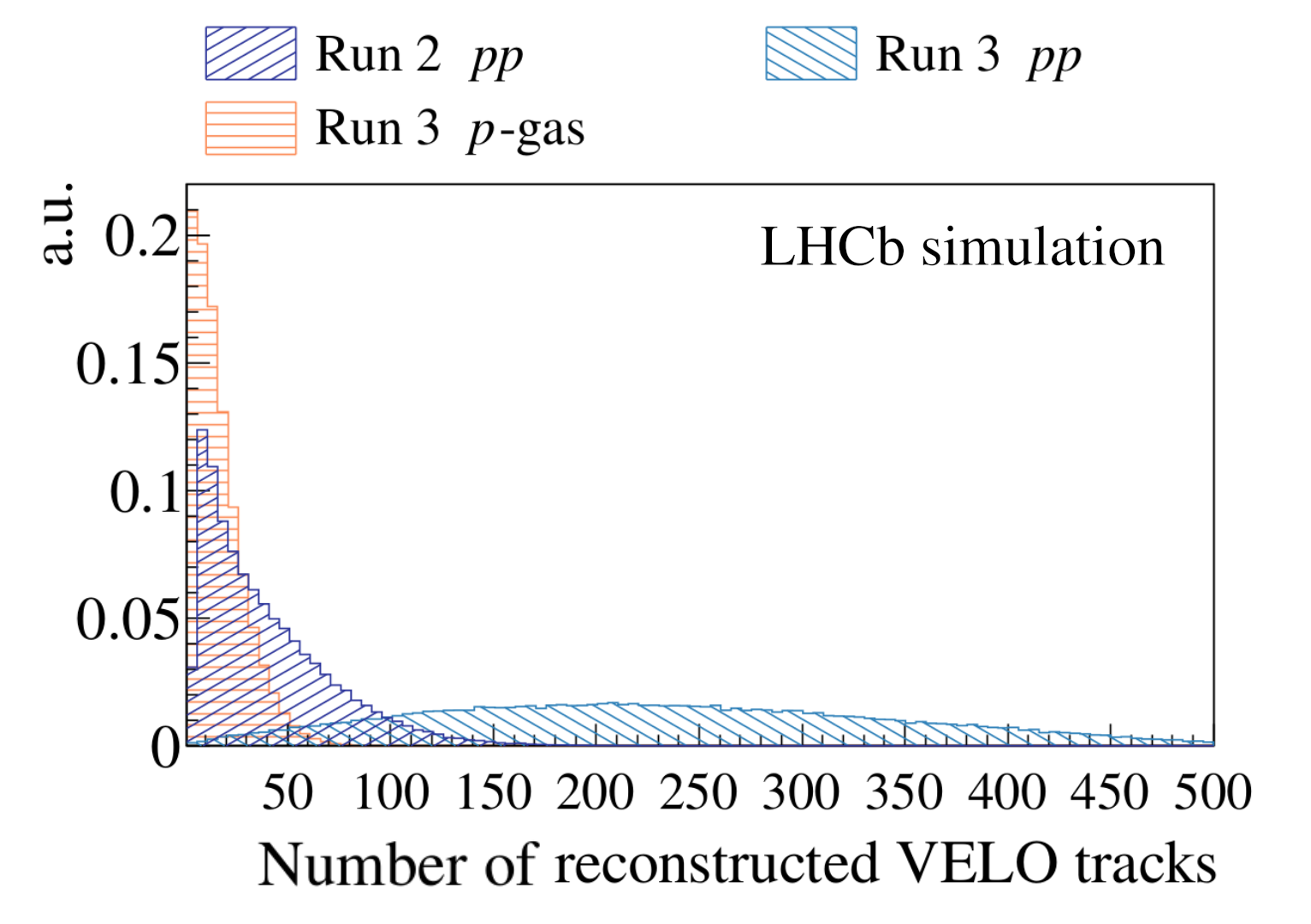}
\caption{A comparison between (left) the average number of visible $pp$ and $p$-gas interactions and between (right) the number of reconstructed \velo tracks in the PVs. For both comparisons, the minimum bias samples are simulated with Run~2 (dark blue histogram) and Run~3 (orange and light blue histograms) beam conditions}
\label{fig:pv_visible}
\end{figure*}

A particularly important part of the \lhcb{} reconstruction is finding the positions of the $pp$ collisions (or primary vertices, PVs), and estimating which charged particle trajectories (tracks) are produced in each PV. In physics analyses, the decay time of long-lived particles is estimated using the positions of the primary and secondary vertices, with the latter referring to the point where a particle decays. Additionally, imposing conditions on whether particles are produced at or away from the PV, when appropriate, serves as a powerful criterion for suppressing background contributions. In the \lhc Run~2 (2015-2018) the data were taken with $\mu = 1.1$. The PV finding algorithms~\cite{Kucharczyk:1756296,Dziurda:2115353} were optimised to this particular running condition by maximising their efficiency and minimising the rate of wrongly reconstructed PVs. A comparison of the average number of visible $pp$ interactions in Run 2 and Run 3 simulated minimum bias events is shown in the left part of Fig.~\ref{fig:pv_visible}. A significant increase of interactions for Run~3 and the coexistence of $pp$ and $p$-gas collisions require a full physics performance reoptimisation. 

To meet all of these challenges a new and intrinsically parallel PV reconstruction algorithm has been developed. In this paper, we describe the key principles of this algorithm, its physics and throughput performance estimated on simulated events. We demonstrate that the algorithm fits within the \lhcb real-time processing resources. \lhcb's heterogeneous processing framework, Allen~\cite{Aaij:2019zbu}, can be used to compile an algorithm for both x86 and GPU architectures. This feature is particularly important when executing \lhcb's GPU processing algorithms on CPU clusters while producing simulated events. However, \lhcb has developed two distinct implementations, each with a logic optimised for the given architecture. Throughout this paper the ``x86 algorithm'' refers to the dedicated x86 implementation, rather than the GPU algorithm compiled for x86 architectures. We compare the x86 and GPU implementations, explaining the different algorithmic choices made to optimise performance on each architecture. 

\section{The \lhcb Upgrade~I detector}
\label{sec:detector}
The \lhcb Upgrade~I detector~\cite{LHCb-TDR-012}\cite{LHCb-DP-2022-002} is a single-arm forward spectrometer covering the pseudorapidity range $2<\eta <5$, designed for the study of particles containing \bquark or \cquark quarks. The \lhcb coordinate system is a right-handed Cartesian system with its origin at the interaction point. The $x$-axis is oriented horizontally towards the outside of the \lhc{} ring, the $y$-axis is pointing upwards with respect to the beamline and the $z$-axis is aligned with the beam direction.

\begin{figure*}
\centering
\includegraphics[width=0.7\linewidth]{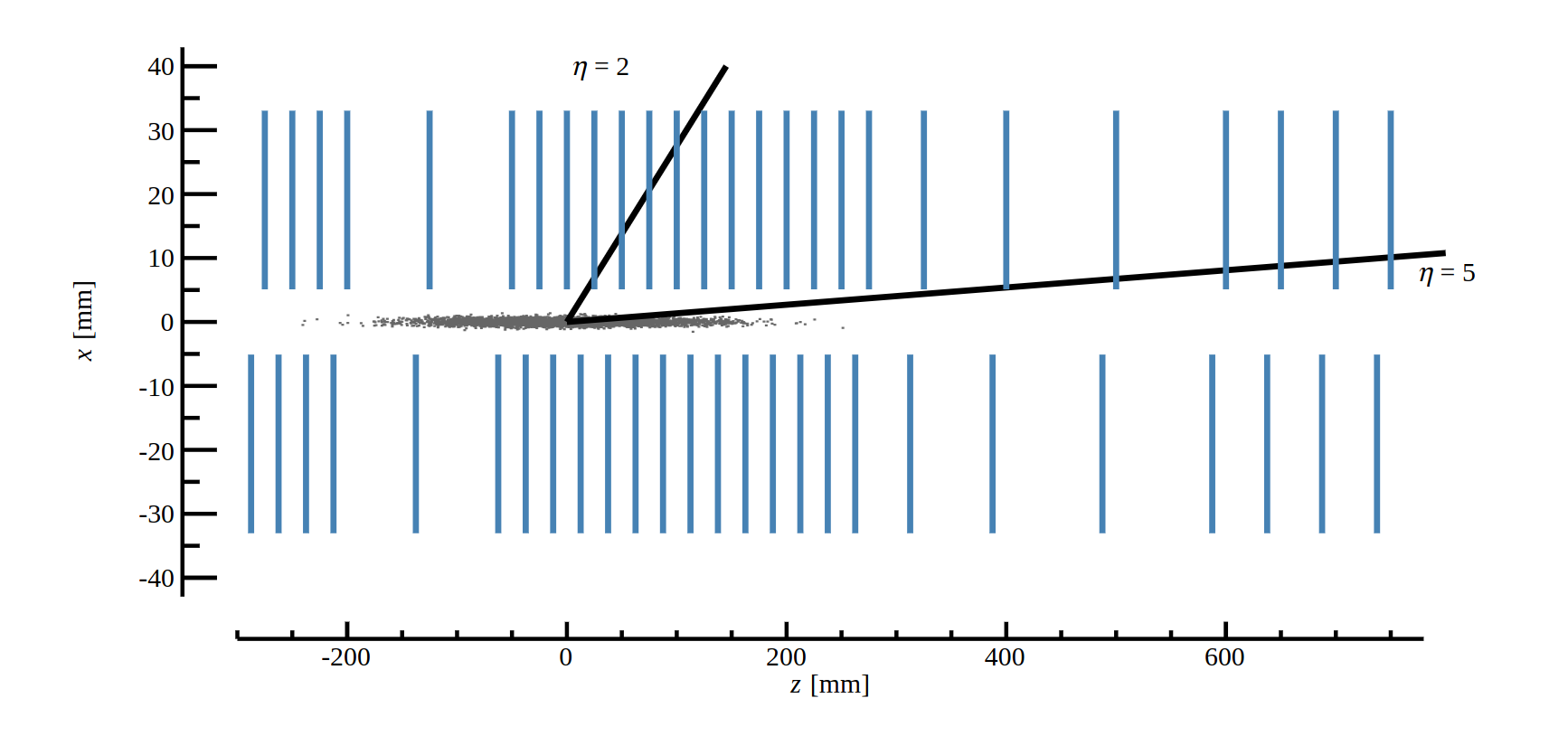}
\caption{The \velo detector geometry, with modules depicted in blue. The \textit{x - z} coordinate system is also shown. Reproduced from Ref.~\cite{LHCb-PUB-2019-008}}
\label{fig:pvinteractionregion}
\end{figure*}

In the context of the PV reconstruction, the most important component is the silicon pixel vertex detector (\velo), which surrounds the interaction region in the forward and backward directions as presented in Fig.~\ref{fig:pvinteractionregion}. The minimal distance of the silicon sensors to the beam is 5.1\mm, in comparison to 8.2\mm for the 2010-2018 \velo. A thin aluminium envelope separates the vacuum around the \lhcb \velo from the \lhc beam vacuum.   

The \lhcb{} PV reconstruction algorithm uses as input tracks reconstructed in the \velo, commonly referred to as \velo tracks. In Run~3 these tracks are reconstructed using the algorithm implemented in x86~\cite{Hennequin:2019itm} and GPU~\cite{VELO_GPU} architectures. Since there is negligible magnetic field in the \velo~\cite{LHCb-TDR-013}, charged particle trajectories are reconstructed as straight lines and their momentum cannot be measured.
Instead, \velo{} track segments are assigned a transverse momentum of 0.4\gevc. The reconstructed pseudorapidity of the track is then used to estimate the momentum. A simplified Kalman filter, which includes the effects of multiple scattering, is performed to estimate the \velo track \textit{x-} and \textit{y-}coordinate positions $(\xtrk,\ytrk)$, direction $(\txtrk\equiv \ud x/\ud z,\tytrk\equiv\ud y/\ud z)$, and their covariance matrix $V$ at a given $z$-coordinate $\ztrk$ near the interaction point. 

A comparison of the number of reconstructed \velo tracks used to form reconstructed PVs for simulated samples with Run 2 and Run 3 beam conditions is shown in the right part of Fig.~\ref{fig:pv_visible}. The essential metric used to determine if a track comes from a primary vertex or from a secondary decay of a long-lived particle is the distance of closest approach of a track to a vertex, called ``impact parameter'' (IP). Depending on the use-case, the \mbox{\textrm{IP} $\chi^2$}, which is the $\chi^2$ difference of a PV reconstructed with and without the track under consideration, is sometimes preferred.

\begin{figure*}
    \centering
   \includegraphics[width=0.49\textwidth]{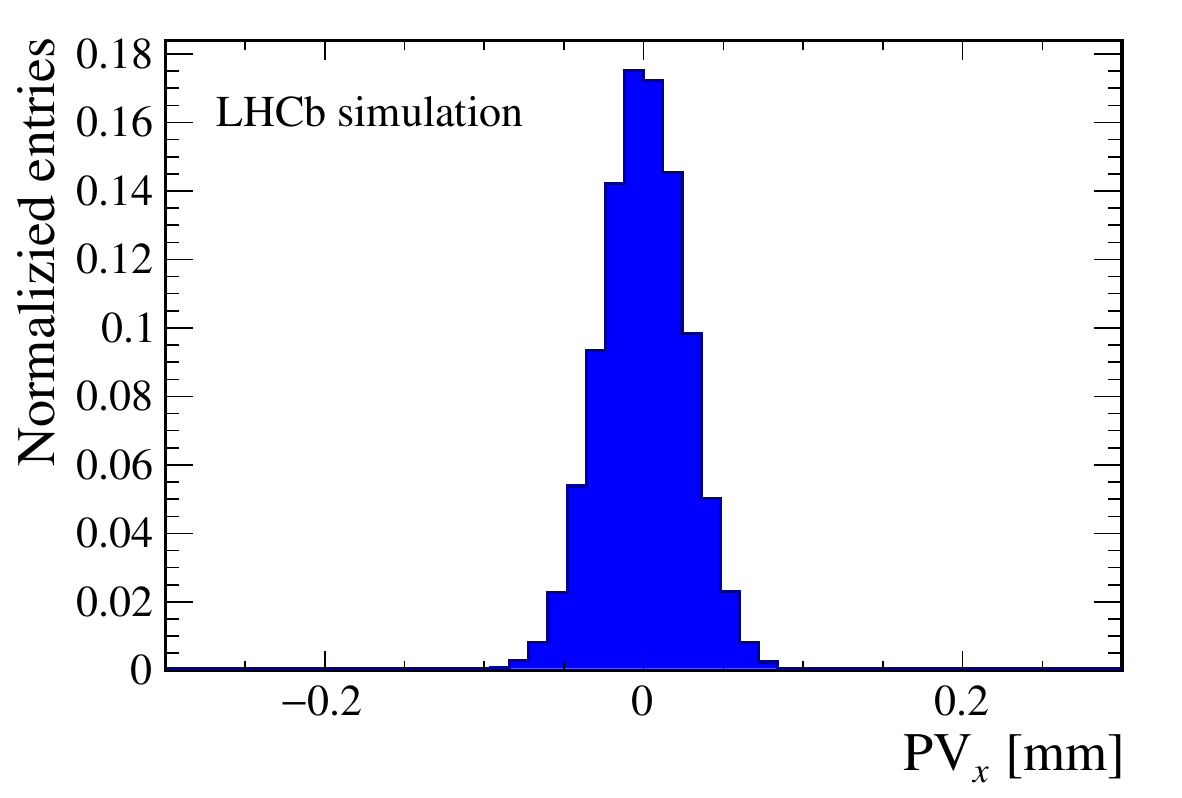}
   \includegraphics[width=0.49\textwidth]{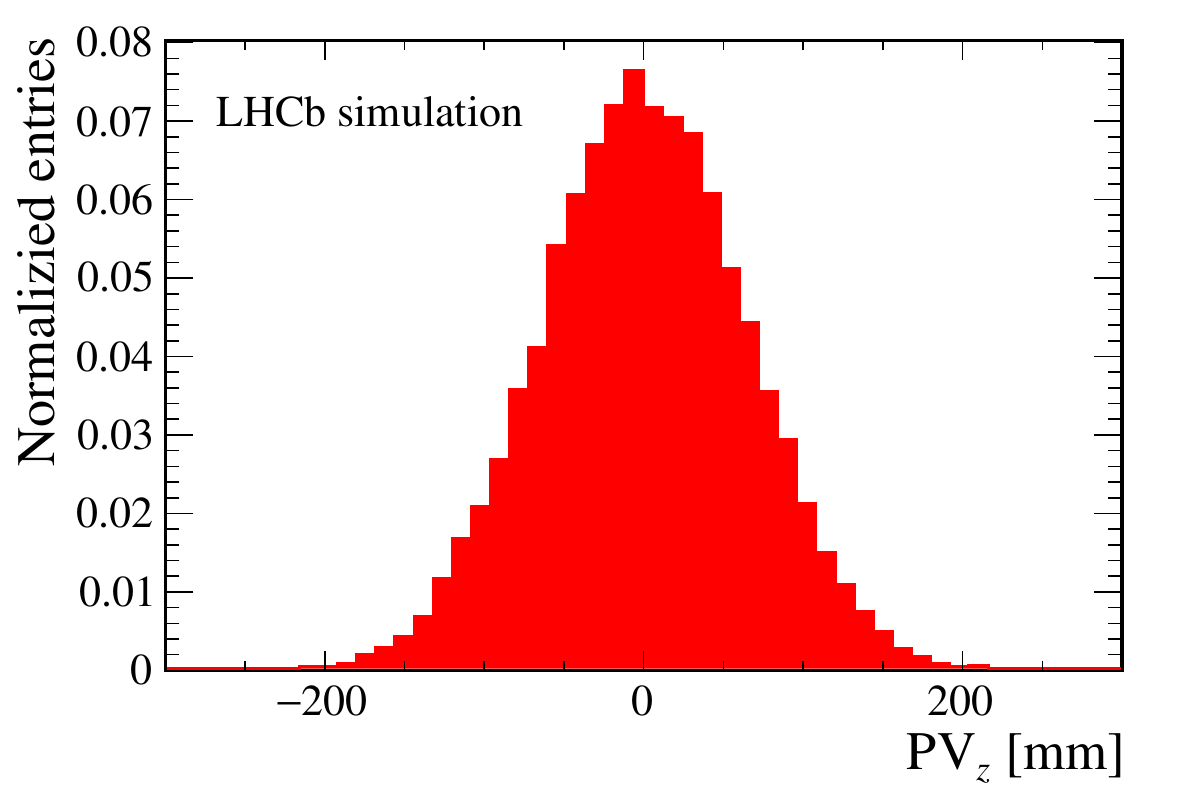}
    \caption{Simulated (left) $x$ and (right) $z$ distribution of $pp$ collisions with Run~3 beam conditions}
    \label{fig:pv_distribution}
\end{figure*}
 \begin{figure*}
 	\centering
 	\includegraphics[width=0.49\linewidth]{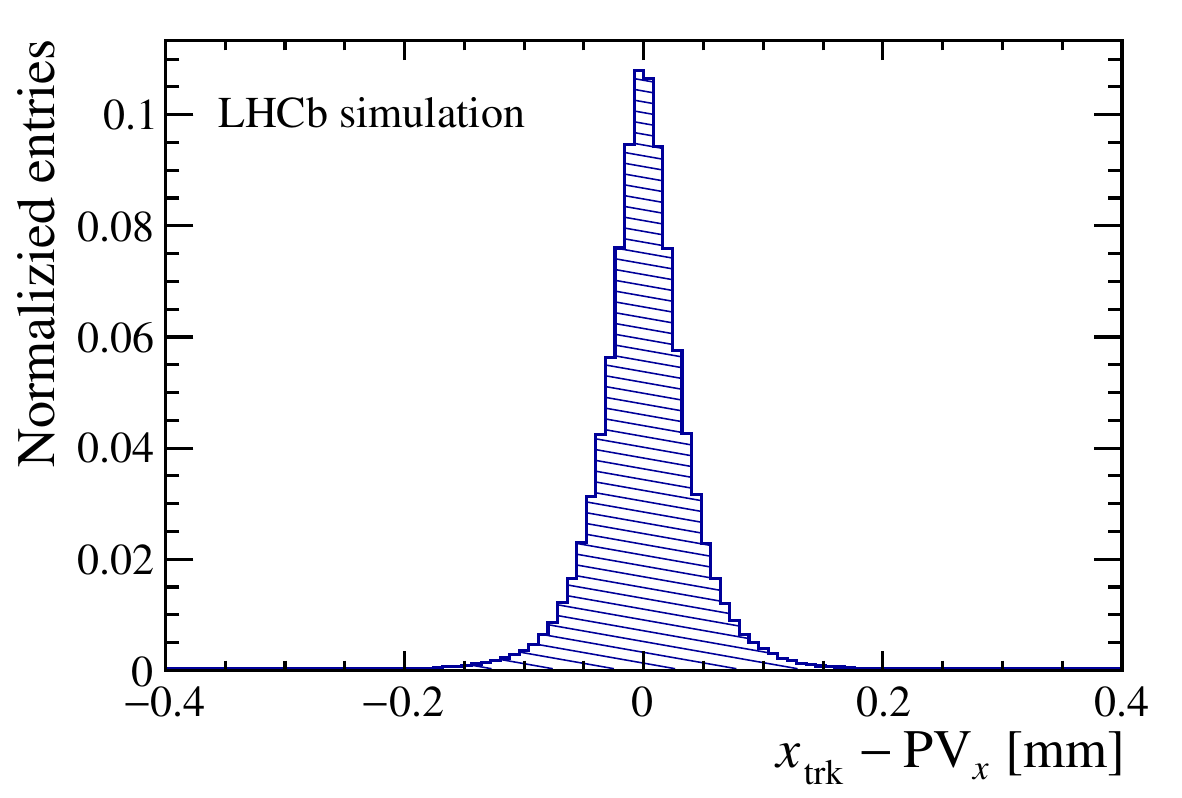}
 	\includegraphics[width=0.49\linewidth]{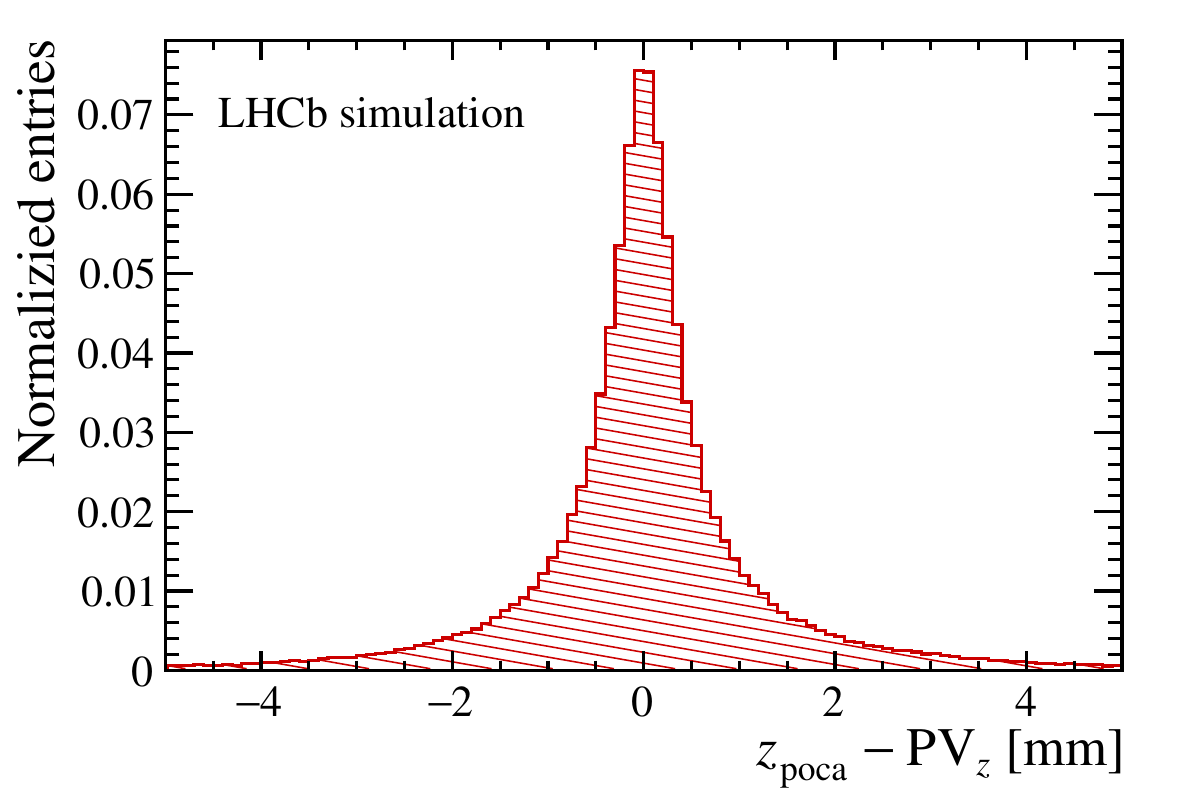}
 	\caption{(Left) distance between the $x$-position of the track, \xtrk, when extrapolated to the $z$-position of its origin PV and the simulated $x$-position of the PV, \pvx. (Right) difference between the $z$-position of the point of closest approach to the beamline of a track originating from the PV, \zpoca and the simulated \pvz. Both distributions are obtained using simulated samples with Run~3 beam conditions}
 	\label{fig:pv_motivation}
 \end{figure*}

\section{Primary vertex finding}
\label{sec:algo}

Primary-vertex-finding algorithms generally consist of a partitioning (or "seeding") step to combine tracks into vertex candidates, followed by an adaptive least squares fit that estimates the vertex position and associated covariance matrix. Traditionally, the partitioning is performed by constructing valid two-prong vertices starting from track pairs~\cite{AreRudi}. These pairs are than combined with the remaining unused tracks in the event to construct multitrack seeds. Because this approach is combinatorial in nature, its complexity grows approximately quadratically with the number of tracks if the number of vertices is larger than one.

The algorithm described here uses a different technique that avoids track-track or track-vertex combinatorics. Its seeding step consists of a one-dimensional histogramming and peak search in the coordinate along the collision axis $z$.  Similar approaches can be found in other LHC experiments~\cite{ATLAS,CMS,ALICE}. In the LHCb experiment, this approach exploits the geometry of the $pp$ interaction region that is spread out in $z$ but narrow in $x$ and $y$\footnote{As the \lhc collision region is symmetric in $x$ and $y$ and the \velo closes by centring on the beam, the $x$ dimension is used to represent both, unless otherwise stated.}, as shown in Fig.~\ref{fig:pv_distribution}.

The algorithm relies on external information of the location of the interaction region. For $pp$ collisions, the interaction region is parametrised as a "beamline" with average transverse position coordinates $\xbeam$ and $\ybeam$ and direction $\txbeam \equiv \ud x/\ud z$ and $\tybeam \equiv \ud y/\ud z$ at $z=0$. During a data-taking period, called ``fill", the \lhc{} beamline does not change on scales relevant to the PV finding algorithm, while it can change between fills. Therefore, a dedicated algorithm (not described further here) is executed in less than a second at the start of every fill to determine the beamline position with a few microns uncertainty. This is subsequently stored in a database and propagated to the PV reconstruction algorithm.
For $p$-gas fixed target collisions, a single beam passing through the target gas is used, resulting in a beamline inclination $\txbeam$, and $\tybeam$ equal to half of the effective crossing angle of the two beamlines in the $pp$ interaction region. The beam inclination effect on the PV reconstruction is significantly more pronounced in $p$-gas collisions, due to the greater lever arm. In contrast, it is negligible for the $pp$ interaction region. The angles are retrieved from the \lhc database and propagated to the PV reconstruction algorithm for each fill.

The main motivation for the one-dimensional histogramming approach can be understood by comparing the distribution of PVs in the $x$ and $z$ coordinates shown in Fig.~\ref{fig:pv_distribution} with the resolution of the track coordinates in simulated events. The left part of Fig.~\ref{fig:pv_motivation} presents the difference between the reconstructed track coordinate $x$ extrapolated to the true $z$ position of the associated PV. The resolution in the track coordinate transverse to the beamline is comparable to the size of the interaction region. Consequently, when assigning individual tracks to PVs, the spread in the transverse coordinate of the PVs is not relevant.

On the other hand, for most tracks the resolution of the coordinate along the beamline is more than sufficient to separate PVs. 
This coordinate is defined as the $z$-coordinate of the point of closest approach to the beamline, given by
\begin{equation}
\begin{split}
  \zpoca  =  \ztrk & + \frac{(\txtrk-\txbeam) (\xbeam - \xtrk)  }
  { (\txtrk-\txbeam)^2 + (\tytrk-\tybeam)^2 }\\ 
  & + \frac{(\tytrk-\tybeam) (\ybeam - \ytrk) }
  { (\txtrk-\txbeam)^2 + (\tytrk-\tybeam)^2 },
  \label{equ:zpocadef}
  \end{split}
\end{equation}
where \xtrk, \ytrk, \ztrk, \txtrk, \tytrk are the track parameters,
while \xbeam, \ybeam, \txbeam, \tybeam are the beamline parameters.
The right part of Fig.~\ref{fig:pv_motivation} shows the difference between the reconstructed $\zpoca$ and the true $z$ coordinate of the associated PV. The distribution is much narrower than the spread of PVs in z-coordinate shown in the right part of Fig.~\ref{fig:pv_distribution}. To find the PVs, the $\zpoca$ values of all track segments in an event are filled into a histogram. Since the tracks originating from a certain PV should have similar values of \zpoca, peaking distributions in the histogram indicate the presence of a PV, roughly at the $z$-position of the peak.

The PV finding algorithm consists of the following steps:
\begin{enumerate}
	\item{\bf \velo tracks extrapolation} to the point of closest approach to the beamline;
	\item{\bf Histogram filling} with the $\zpoca$ value of each track;
	\item{\bf Peak search} in the histogram;
	\item{\bf Track association} to the identified peaks;
	\item{\bf Vertex fit} using the assigned tracks.
\end{enumerate}
The major difference between the two hardware architectures occurs at the track association stage and propagates to the fitting procedure.

\subsection{\velo track preparation}

A simplified Kalman filter, which includes the effects of multiple scattering, is performed to estimate the track parameters which are subsequently used to compute $\zpoca$ according to Eq.~\ref{equ:zpocadef}. The uncertainty in $\zpoca$ strongly depends on the track slope and the distance to the first hit on the track. For the performance of the histogramming method, it is important to exploit the variation in this uncertainty.

The estimated \zpoca{} uncertainty can be computed from the state covariance matrix. Given the track parameter covariance matrix $V$ at position $\ztrk$, the covariance matrix for the transverse coordinates extrapolated to position $z$ is given by
\newcommand{\dz}{{\Delta z}}
\begin{equation}
\begin{split}
  V_{x,y} (z) & \; = \; 
  \begin{pmatrix}
    V_{xx} & V_{xy} \\
    V_{xy}  & V_{yy}  \\
  \end{pmatrix} \\
  & \; + \; 
  \begin{pmatrix}
   2 \dz \, V_{x t_x}  &  \dz \, V_{x t_y} \\
    \dz \, V_{x t_y}  &  2 \dz \, V_{y t_y}  \\
  \end{pmatrix} \\
 & \; + \; 
  \begin{pmatrix}
     \dz^2 \, V_{t_x t_x} &  \dz^2 \, V_{t_y t_x}\\
     \dz^2 \, V_{t_y t_x}  &  \dz^2 \, V_{t_y t_y} \\
  \end{pmatrix},
  \label{equ:Vxy}
  \end{split}
\end{equation}
with $\dz = z - \ztrk$ and where $V_{ij}$
indicates  the \textit{ij} element of the covariance matrix $V$. 

\newcommand{\sigmaz}{\ensuremath{\sigma_\text{poca}}}

In the simplified \velo{} track fit, the multiple scattering is treated independently in $x$ and $y$. As a consequence, the estimated $x$ and $y$ track coordinate uncertainties are also assigned identical magnitudes and treated as uncorrelated. Under these assumptions, linear error propagation of Eq.~\ref{equ:zpocadef} results in 
\begin{equation}
    \sigmaz = \sqrt{V_{xx} / \left( \left(\txtrk - \txbeam\right)^2 + \left(\tytrk - \tybeam\right)^2 \right)}.
\end{equation}
The uncertainty is approximately inversely proportional to the track slope.

Together with the track parameters, this constitutes all the necessary inputs for later algorithm stages. The track parameters and the inverted covariance matrix $W$ of each track are computed once and used throughout the rest of the algorithm. This approach has been found to have a negligible impact on the performance of the algorithm, but improves its throughput.

\subsection{Histogram filling}

In the next step, the $\zpoca$ values of the tracks are used to fill a histogram. The histogram boundaries are $[-550,300]$\mm, spanning both the main $pp$ interaction region ($z \approx 0$\mm) and the SMOG2 gas cell ($z\approx-450$\mm). The bin size, $dz$, is chosen to be $0.25$\mm.  In the peak search, the minimum distance between two peaks is two times the bin size. Consequently, the bin size determines the minimal PV separation. In practice, in the $pp$-interaction region the algorithm cannot separate vertices that are closer than about 2\mm.

\begin{figure*}
  \centering
  \includegraphics[width=0.49\linewidth]{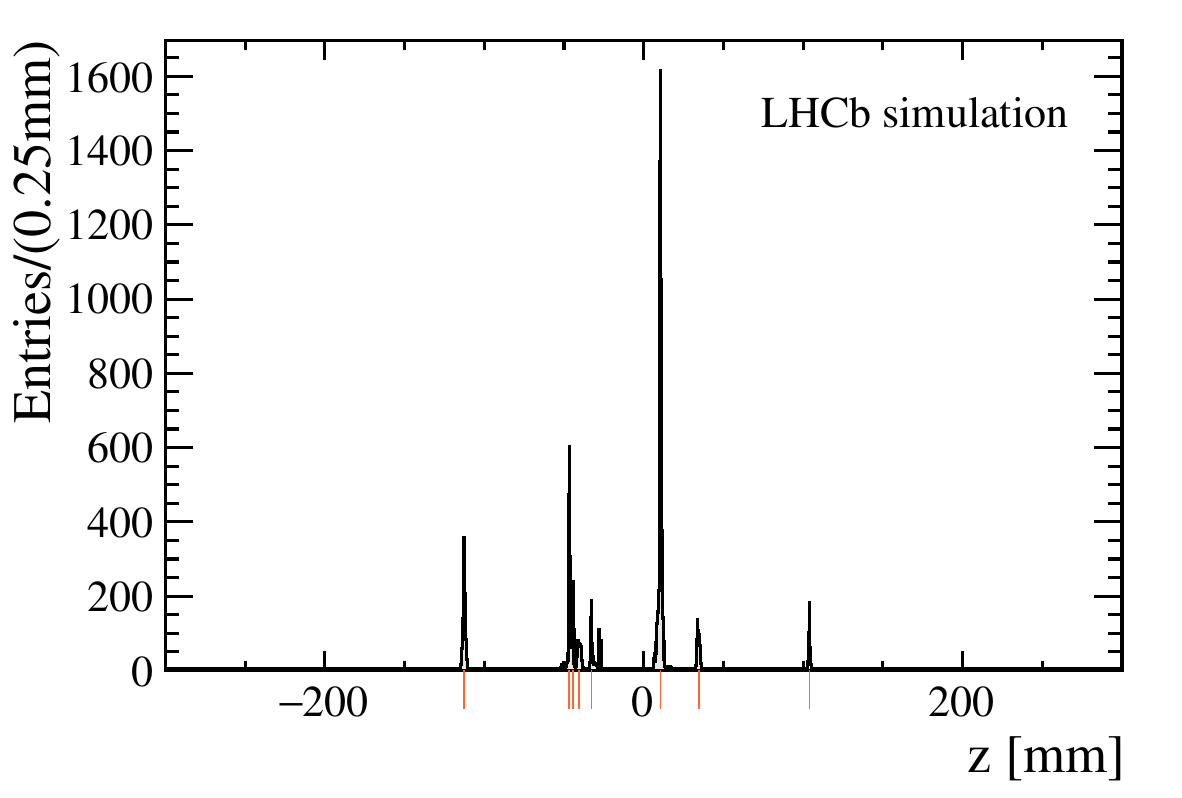}
    \includegraphics[width=0.49\linewidth]{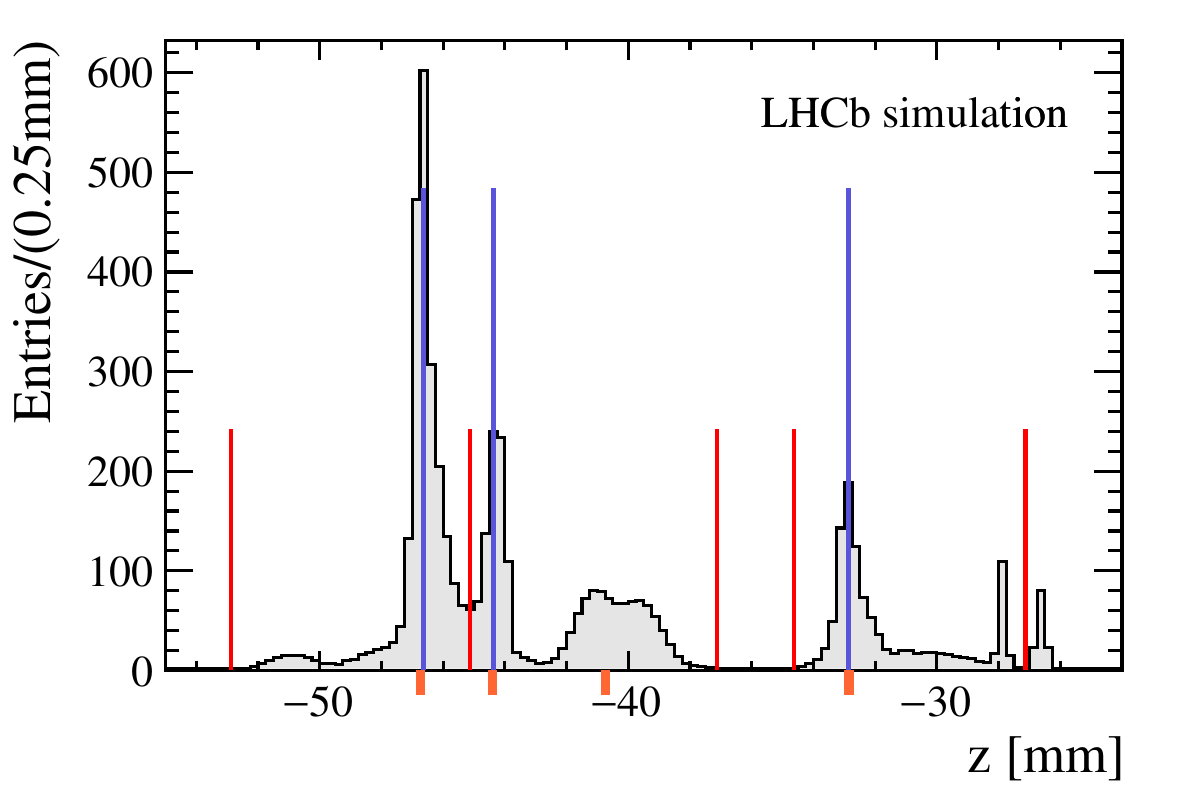}
  \caption{Typical histogram filled by the PV reconstruction algorithm with the \zpoca values, following the method explained in the text. The orange markers indicate the position of the simulated reconstructible vertices, defined in Sec.~\ref{sec:physics}. The right plot is a zoom of the top distribution on some of the identified peaks. These are shown as blue vertical lines and the borders of the peaks as vertical red lines}
  \label{fig:histofullrange}
\end{figure*}

To reduce effects due the choice of the bin size and the uncertainty of the track extrapolation, a single track may contribute to multiple bins. The contribution to bin $i$ with bin boundaries $[z_{i,\text{min}},z_{i,\text{max}}]$ is calculated by the integral over the bin of a Gaussian distribution with mean $\zpoca$ and standard deviation $\sigmaz$. 
With the given histogram range, the contributions $\omega_i$ for a single track add up to unity. 
To reduce computation costs, the integrals are only performed on a finite number of bins neighbouring the central bin that contains $\zpoca$. Furthermore,
the maximal $\sigmaz$ for adding a track to the histogram is set to $1.5$\mm for the $pp$ interaction region, and $10$\mm for the gas-cell region. An example histogram covering part of the $z$-range is shown in left Fig.~\ref{fig:histofullrange}, where peaking structures likely corresponding to PVs for $pp$ collisions can be seen. Positions of the reconstructible simulated PVs, defined as in Sec.~\ref{sec:physics}, are also indicated by the orange markers.

The integrals of the Gaussian kernel are relatively expensive to compute. Therefore, we have developed two types of approximations. The x86 implementation works with a finite set of template histograms, selected based on the value of $\sigmaz$.
In the GPU implementation, we rely on a cubic polynomial approximation of the cumulative density of the Gaussian distribution.

\subsection{Peak search}
After filling the histogram, a peak search is performed. In a first step, ``proto-clusters" are identified as regions of subsequent bins with content above a threshold. Since two PVs might be very close in $z$, such that there are no bins below the threshold separating their proto-clusters, a dip search is then performed to be able to split them into seed clusters. First, all significant minima and maxima in the range of histogram bins of a proto-cluster are identified. The proto-cluster is then split into seed clusters at minima which have two neighbouring maxima. The splitting is only done if the track integral of the resulting seed cluster is above a threshold, which effectively means that enough tracks will contribute to the vertex fit.

The logic of the splitting of proto-clusters differs between the x86 and GPU implementations. While the GPU implementation iterates through the minima ordered in $z$ as potential splitting points, the CPU implementation considers the smallest minimum between the two largest maxima and does the splitting recursively.

In the second step, the $z$-position of a cluster is computed as
\begin{equation}
    z_\text{seed} \; = \; z_{i} + \delta \times dz,
\end{equation}  
where $i$ is the bin with the maximum content in the cluster and $z_i$ is the midpoint of this bin. The correction $\delta$ is computed from the bin content $N_i$ and that of the neighbouring bins as
\begin{equation}
    \delta = \frac{1}{2} \frac{N_{i+1} - N_{i-1}}{2N_{i} - N_{i+1} - N_{i-1}}.
\end{equation}
As $N_{i\pm1} < N_i$, this correction is in the range $[\text{-}1/2, 1/2]$ and $\delta \rightarrow \pm 1/2$ in the limit $N_{i\pm 1} \rightarrow N_i$. By construction, the produced clusters are ordered in $z$. Once the $z$-position of the seed is computed, its $xy$ position is evaluated using the known beamline.

The right part of Fig.~\ref{fig:histofullrange} shows the result of the seed reconstruction, zoomed in on a subset of the identified peaks, for a typical event. The edges of each cluster are indicated in red, while the identified peaks are denoted as blue lines. In this event, the simulated PV close to $z=-40\mm$ is not reconstructed by the algorithm.

\subsection{Tracks association}
   
A different tracks-to-PV association strategy is chosen for the x86 and the GPU algorithm implementations. While in the former a track is only associated to one PV, in the latter each track can contribute to multiple PVs with different weights. 

The CPU time consumption of the x86 algorithm is dominated by the vertex fit. The time is proportional to the number of tracks per vertex, the number of vertices and the number of iterations of the vertex fit. Therefore, to minimise the CPU time, every track is associated to a single vertex and  the track-to-vertex assignment is stored in a look-up table. First, every bin in the \zpoca{} histogram is assigned an index corresponding to the closest cluster. This is performed by first determining partitioning points, which are defined as the bin in between of the upper bound of one cluster and the lower bound of the next cluster. Subsequently, the bins in between the partitioning points are assigned: this requires effectively a single loop over all bins. Once the histogram bins have been assigned, for every track the index of its vertex is found by computing its \zpoca{} bin, which requires a single loop over all tracks. After the partitioning is performed, the number of tracks per cluster is determined: In the rare case that this is smaller than a threshold (respectively 3 for the $pp$ interaction region and 2 for the SMOG2 region), the seed cluster is removed form the list of clusters and the procedure is repeated. The advantage of this approach is that it does not require track-vertex combinatorics, nor any comparison operations. 

The iterative procedure is less suitable for the GPU implementation as tracks have to be re-distributed if a vertex candidate has too few associated tracks. This creates dependencies between the vertex candidates and limits the parallelism that can be achieved. In contrast, the computation of each track's PV weights can be fully parallelised. The implicit track association is therefore chosen, allowing tracks to contribute to more PVs during the vertex-fit procedure, as described in the next section.

\subsection{Vertex fit}

\label{sec:vertex_fit}

To improve the resolution of the vertex position and compute the associated covariance matrix, each seed is fitted with a least squares method. The general vertex-fit procedure is first described for the case of the explicit track association, and subsequently the modifications needed for the implicit case are discussed. 

The vertex fit minimises a $\chi^2$ defined as
\begin{equation}
  \chi^2 \; = \; \sum_{\text{tracks $i$}} w_i \: \chi^2_i
\end{equation}
with respect to the vertex position. In this expression, $w_i$ is the weight of the track $i$, discussed later, and $\chi^2_i$ is the contribution of track $i$ to the vertex. The latter can be written as
\begin{equation}
  \chi^2_i \; = \; r_i^{T} \: V_i^{-1} \: r_i,
  \label{equ:chi2_definition}
\end{equation}
where $r_i$ is the residual of the track $i$ and $V_i$ is the state covariance matrix. In the vertex fit where tracks are explicitly associated to the vertex, the parameters of the fit are the vertex position $\vec{x}_\text{vtx}$  and the outgoing momentum vectors (or eventually direction vectors) $\vec{p}_i$ of all of the tracks. The five-component residual can then be expressed as
\begin{equation}
  r_i \; = \; m_i \: - \: h_i( \vec{x}_\text{vtx}, \vec{p}_i ), 
\end{equation}
where $m_i$ are the (five) track parameters and $h_i$ is usually called the measurement model. In this case $V_i$ is the covariance matrix of the track parameters $m_i$. An efficient implementation of the ordinary \lhcb vertex fit can be found in~\cite{Amoraal:2012qn}.

The minimisation of the $\chi^2$ to the momentum vectors of the tracks makes the vertex fit nonlinear. Therefore, to minimise CPU costs, it is chosen not to minimise with respect to these parameters for the PV fit.  In the first iteration of a vertex fit, the momentum parameters are usually initialised with the measured momentum parameters of the track. As a result only two components of the residual are nonzero. These can be chosen as
\begin{equation}
  r_i \; = \;
  \begin{pmatrix}
    \xtrki + (\zvtx - \ztrki) \cdot \txtrki - \xvtx \\
    \ytrki + (\zvtx - \ztrki) \cdot \tytrki - \yvtx
  \end{pmatrix}.
\end{equation}
The corresponding $2\times2$ covariance matrix $V_i$ is given by~Eq.~\ref{equ:Vxy} with $\dz = \zvtx - \ztrki$. The disadvantage of this choice for the residual $r_i$ is that the covariance matrix $V_i$ of the residual depends on the vertex position. To minimise CPU costs, the covariance matrices $V_i$ are evaluated and inverted only once, using the vertex position of the seed, discussed above. Since the initial $z$ position is close to the final fitted $z$ position for the majority of vertex seeds, this choice has negligible impact on the performance. Furthermore, because the $x$ and $y$ projections of the \velo{} track fit are independent, the matrix $V_i$ is diagonal, which can be exploited to simplify the expressions in the actual implementation. 

The vertex $\chi^2$ is minimised with the Newton--Raphson method. The first and second derivatives of the $\chi^2$ are computed with respect to the vertex parameters \mbox{$\alpha\equiv(\xvtx,\yvtx,\zvtx)$} and are given by
\begin{equation}
  \frac{\partial \chi^2}{\partial\alpha} \; = \; 2 \sum_{\text{tracks $i$}} w_i H_i^T V_i^{-1} r_i,
\end{equation}
\begin{equation}
  \frac{\partial^2 \chi^2}{\partial\alpha^2} \; = \; 2 \sum_{\text{tracks $i$}} w_i H_i^T V_i^{-1} H_i,
\end{equation}
where $H_i$ is the derivative (or projection) matrix
\begin{equation}
  H_i \; \equiv \; \frac{\partial r_i}{\partial\alpha}
  \; = \; \begin{pmatrix}
    -1 & 0 & \txtrk \\
    0  & -1 & \tytrk
  \end{pmatrix}.
\end{equation}
Note that in these expressions we explicitly ignore the dependence of $V_i$ (and eventually $w_i$) on $\alpha$. Given an initial vertex position $\alpha_0$, the solution that minimises the $\chi^2$ is now given by 
\begin{equation}
  \alpha_1 \; =\; \alpha_0 - \left( \left.\frac{\partial^2 \chi^2}{\partial\alpha^2}\right|_{\alpha_0} \right)^{-1} \left.\frac{\partial \chi^2}{\partial\alpha}\right|_{\alpha_0},
\end{equation}
with the derivatives evaluated using $\alpha=\alpha_0$. The estimated covariance matrix for $\alpha_1$ is
\begin{equation}
  C \; = \; \left( \frac{1}{2} \: \left.\frac{\partial^2 \chi^2}{\partial\alpha^2}\right|_{\alpha_0} \right)^{-1}. 
\end{equation}
The expected $\chi^2$ of the new solution can be computed as
\begin{equation}
  \chi^2_1 \; = \; \chi^2_0 + \Delta\chi^2,
\end{equation}
with the expected change in $\chi^2$ given by
\begin{align}
  \Delta\chi^2 & = \;
  (\alpha_1 - \alpha_0) \left.\frac{\partial \chi^2}{\partial\alpha}\right|_{\alpha_0} +
  \frac{1}{2} (\alpha_1 - \alpha_0)^2 \left.\frac{\partial^2 \chi^2}{\partial\alpha^2}\right|_{\alpha_0} \\
 & \; = \;
  \frac{1}{2} (\alpha_1 - \alpha_0) \left.\frac{\partial \chi^2}{\partial\alpha}\right|_{\alpha_0}.
  \label{equ:deltachisq}
\end{align}

If not for the presence of the weights $w_i$, the solution would be exact and no fit would be required. In practice, the weights discussed below make the fit strongly nonlinear. Therefore the vertex fit requires multiple iterations, with the residuals and derivatives for the next iteration evaluated using the last best estimate of the vertex position. A convergence criterion is chosen based on the $\Delta\chi^2$ in Eq.~\ref{equ:deltachisq} and the observed change in $\zvtx$.  The motivation not to rely on $\Delta\chi^2$ only is that because of possible large variations in the weights, the $\Delta\chi^2$ evaluated using Eq.~\ref{equ:deltachisq} is not always a good estimate of the actual change in $\chi^2$. For the x86 implementation the vertex fit is considered converged if $|\Delta\chi^2|<0.01$ and $|\Delta\zvtx|<1$~\mum. For the GPU implementation, the criterion is $|\Delta\zvtx|<0.5$~\mum{} with no requirement on  $\Delta\chi^2$. To limit the CPU costs due to poorly converging fits, the maximum number of iterations is set to ten. The fits typically converge in three to seven iterations.

To reduce the effect of tracks that are mistakenly assigned to the vertex, track contributions to the vertex $\chi^2$ are weighted with a weight $w_i$ that is a function of $\chi_i$. In the x86 implementation these weights are chosen according to Tukey's bi-square function~\cite{beaton1974,hampel1986},
\begin{equation}
  w_i = \begin{cases}
    \left( 1 - \chi^2_i/\chisqmax \right)^2 & \text{for $\chi^2_i < \chisqmax$} \\
    0 & \text{for $\chi^2_i \ge \chisqmax$},
  \end{cases}
  \label{eq:tukey}
\end{equation}
where $\chisqmax$ is a cut-off value set to 12, that was optimised by considering both the effect of the tails on the resolution and the impact on the efficiency of low-multiplicity vertices.

\begin{figure}
  \centerline{\includegraphics[width=\linewidth]{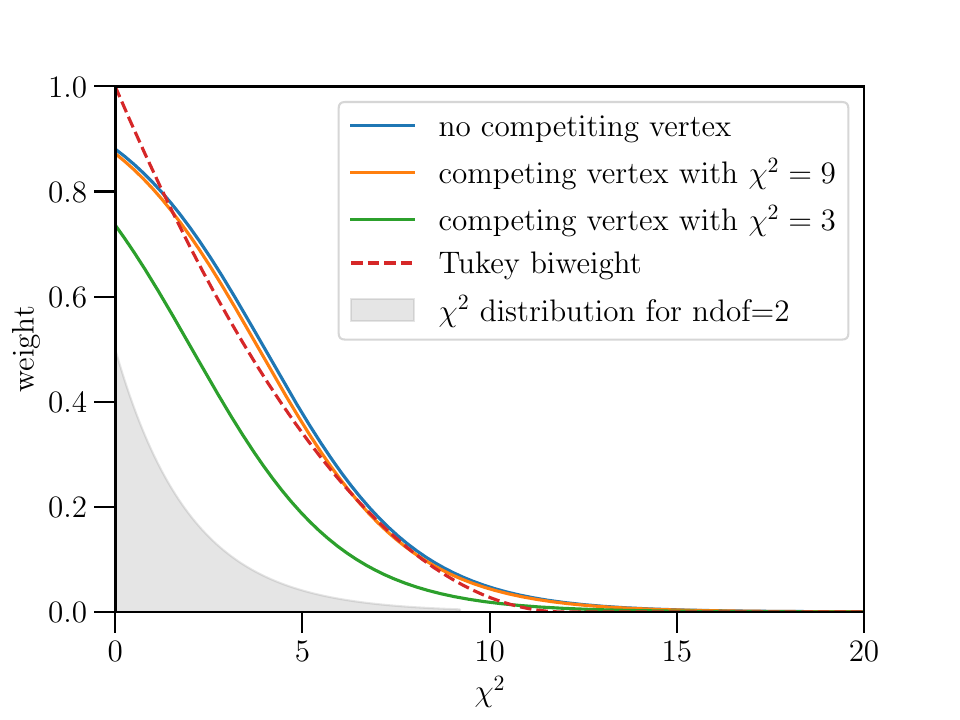}}
  \caption{Multivertex track weights as function of $\chi^2$ for a vertex candidate with no competing vertices (blue), a competing vertex with $\chi^2$=9 (orange) and $\chi^2=3$ (green), all with $\chisqmax=4$. The Tukey biweights with a $\chi^2_{max}=12$ are shown as red dashed line. The grey filled area shows the $\chi^2$ distribution for two degrees of freedom}
  \label{fig:toymultiadaptiveweight}
\end{figure}

In the GPU implementation, all PVs are fitted simultaneously as inspired by multivertex fitter algorithms~\cite{Fruhwirth:2007hz} and every track is implicitly associated to every vertex.
The weights are chosen such that they not only depend on the $\chi^2_i$ of a track with respect to the closest vertex candidate $j$, but also on the other vertex candidates~\cite{Fruhwirth:2007hz} 
\begin{equation}
w_{ij} = \frac{\exp(-\chi^2_{ij}/2)}{\exp(-\chisqmax/2)+ \sum_{k} \exp(-\chi^2_{ik}/2)},
\end{equation}
where $\chisqmax$ is a cut-off value and the $\chi^2_{ik}$ is the $\chi^2$ of track $i$ relative to vertex $k$, evaluated as in Eq.~\ref{equ:chi2_definition}.
This means that a track close to two vertex candidates contributes to both but with a smaller weight than would be the case if it had been explicitly assigned to a PV. Figure~\ref{fig:toymultiadaptiveweight} illustrates $w_{ij}$ as function of $\chi^2$ in the case where no other competing vertices are in proximity and in the case where there are other PVs with $\chi^2=9$ and $\chi^2=3$ with respect to that track. The parameter controlling the steepness of the curve is $\chi^2_{max}$. The smaller its value, the faster the weight falls to zero with increasing $\chi^2$. It is not a strict cut-off like in the case of the Tukey weight introduced earlier, but can be understood as the $\chi^2$ value at which the weight is equal to $0.5$ in the case of no other competing vertices. If there are no other consistent vertices nearby, this weight function becomes similar to the Tukey weight. To keep the fit of a certain vertex candidate independent of the other vertex fits in the event, the weight is always calculated using the initial position of the other vertices. To reduce the number of duplicate PVs, where one collision is reconstructed as two separate PVs, an additional step searches for PVs within close proximity (by default requiring the ratio between the difference of the two \pvz position and the sum of the two \textit{z} variances to be below 25) and rejects the one with fewer associated tracks.

After all vertex positions have been obtained, a final selection is applied to reduce the contamination by secondary decay vertices of long-lived particles to an acceptable, below 2\%, level. Vertices in the $pp$ interaction region that are not within 0.3\mm distance from the beamline are discarded.

\section{Physics performance}
\label{sec:physics}
The physics performance is determined with minimum bias samples produced under Run~3 conditions and the full \lhcb simulation~\cite{LHCb-PROC-2010-056}\cite{LHCb-PROC-2011-006}\footnote{The PV distribution is generated with the mean position at ($x$,$y$,$z$)=(1.092, 0.472, -0.0115)\mm. The performance studies are conducted on 2.5 million events.}. The same events are used for the evaluation of x86 and GPU performance, allowing for a direct comparison of the algorithms. In addition, the Run~2 PV reconstruction algorithm, referred to as Run2-like hereafter, has been optimised for Run~3 conditions~\cite{DeCian:2309972} and is compared with the x86 and GPU implementations described in this paper. The Allen framework allows to compile algorithms developed for GPU architectures also on x86 architectures. A detailed comparison of the physics performance and reproducibility between different architectures is out of the scope of this paper, but in general agreement at the permille-level is observed.

A reconstructible PV (PV$^{\rm{MC}}_{\rm{rcible}}$) is defined as an inelastic interaction which produces at least four reconstructed \velo tracks. This criterion is lowered to three reconstructed \velo tracks for $p$-gas collisions due to their lower average PV-track multiplicity. A reconstructed vertex (PV$^{\rm{REC}}$) is matched to a simulated reconstructible PV if the distance between the simulated and reconstructed \textit{z}-coordinate of PV is lower than five times its reconstruction uncertainty. If a simulated PV is matched to more than one reconstructed PV, only the closest match is retained. The reconstructed and matched primary vertices (PV$^{\rm{REC}}_{\rm{matched}}$) are then selected to measure the following figures of merit.  

\begin{enumerate}
\item{\textbf{Efficiency},} defined as the ratio of reconstructed and matched PVs to the total number of reconstructible PVs in the simulation
\begin{equation}
	\epsilon = \frac{\text{\# PV}^\text{REC}_\text{matched}}{\text{\# PV}^\text{MC}_\text{rcible}}.
\end{equation}
A low efficiency would result in some prompt tracks being identified as originating from decays of long-lived particles, increasing background for the real-time processing and physics analyses.
\item{\textbf{Fake rate},} defined as the ratio of reconstructed, but not matched PVs to the total number of reconstructed PVs
     \begin{equation}
        f = \frac{\text{\# PV}^\text{REC} - \text{\# PV}^\text{REC}_\text{matched}}{\text{\# PV}^\text{REC}}.
        \end{equation}
      Most fake PVs are secondary vertices from decays of long-lived particles. Thus, a high fake rate would reduce the signal efficiency for physics analyses.
\item{\textbf{Position resolution},} defined as the standard deviation of the distribution of the difference between a reconstructed and its matched simulated PV position. The PV resolution is an important component of the decay-time resolution for long-lived particles and of the track IP resolution. For the latter, it is particularly important for high-momentum tracks, which undergo little multiple scattering and whose IP resolution is therefore dominated by the PV resolution itself.
\item{\textbf{Pull},} defined as the ratio between the position resolution and the reconstruction uncertainty. An optimal pull distribution has zero mean and unit width, while deviations hint at biases in the PV position reconstruction or a not accurately estimated covariance matrix.
\end{enumerate}

The algorithm performance is studied as a function of different quantities such as the PV \textit{z} position, the number of particles associated to the primary vertex called PV multiplicity, and for different vertex categories. A PV is defined as \texttt{close} if any reconstructible neighbouring PV is closer than 10\mm. Otherwise, the PV is labelled as \texttt{isolated}. For the purpose of performance categorisation, PVs are sorted from highest to lowest \velo track multiplicity (\texttt{1st}, \texttt{2nd}, \texttt{3rd}, ...). Finally, PVs which produce particles containing quark species (\texttt{beauty}, \texttt{charm}, \texttt{strange}, \texttt{other}) are benchmarked separately. The PV multiplicity varies across categories, averaging 68, 63, and 37 associated particles for \texttt{beauty}, \texttt{charm}, and \texttt{strange} PVs, respectively. In contrast, the \texttt{other} category has a significantly lower average of just 7 associated particles, resulting in a much lower reconstruction efficiency.

\subsection{Performance for \textit{pp} collisions}
\begin{table}
\caption{Primary-vertex-reconstruction efficiency for x86, GPU and Run2-like implementations. Different primary vertex categories for the $pp$ conditions are listed as described in the text. All numbers are given in percentages}
\label{tab:eff}
\begin{tabular}{llll}
\noalign{\smallskip}\hline\noalign{\smallskip}
Category & x86 & GPU & Run2-like \\ 
\hline\noalign{\smallskip}
All               & 93.3 & 93.7 & 91.3 \\
\texttt{beauty}   & 98.1 & 98.4  & 98.2 \\ 
\texttt{charm}    & 98.0& 98.3  & 98.3  \\
\texttt{strange}  & 93.5& 93.9  & 91.5   \\
\texttt{other}    & 63.3& 63.5 & 48.9  \\
\texttt{isolated}    & 97.6 & 97.6 & 95.4\\
\texttt{close} & 89.0 & 89.7 & 87.1  \\
\texttt{1st}      & 99.5 & 99.5 & 99.5 \\
\texttt{3rd}      & 96.2& 96.5 & 95.4   \\
\texttt{5th}      & 90.5 & 91.1 & 87.5\\
\noalign{\smallskip}\hline
\end{tabular}
\end{table}

\begin{figure*}
	\centering
	\includegraphics[width=0.49\linewidth]{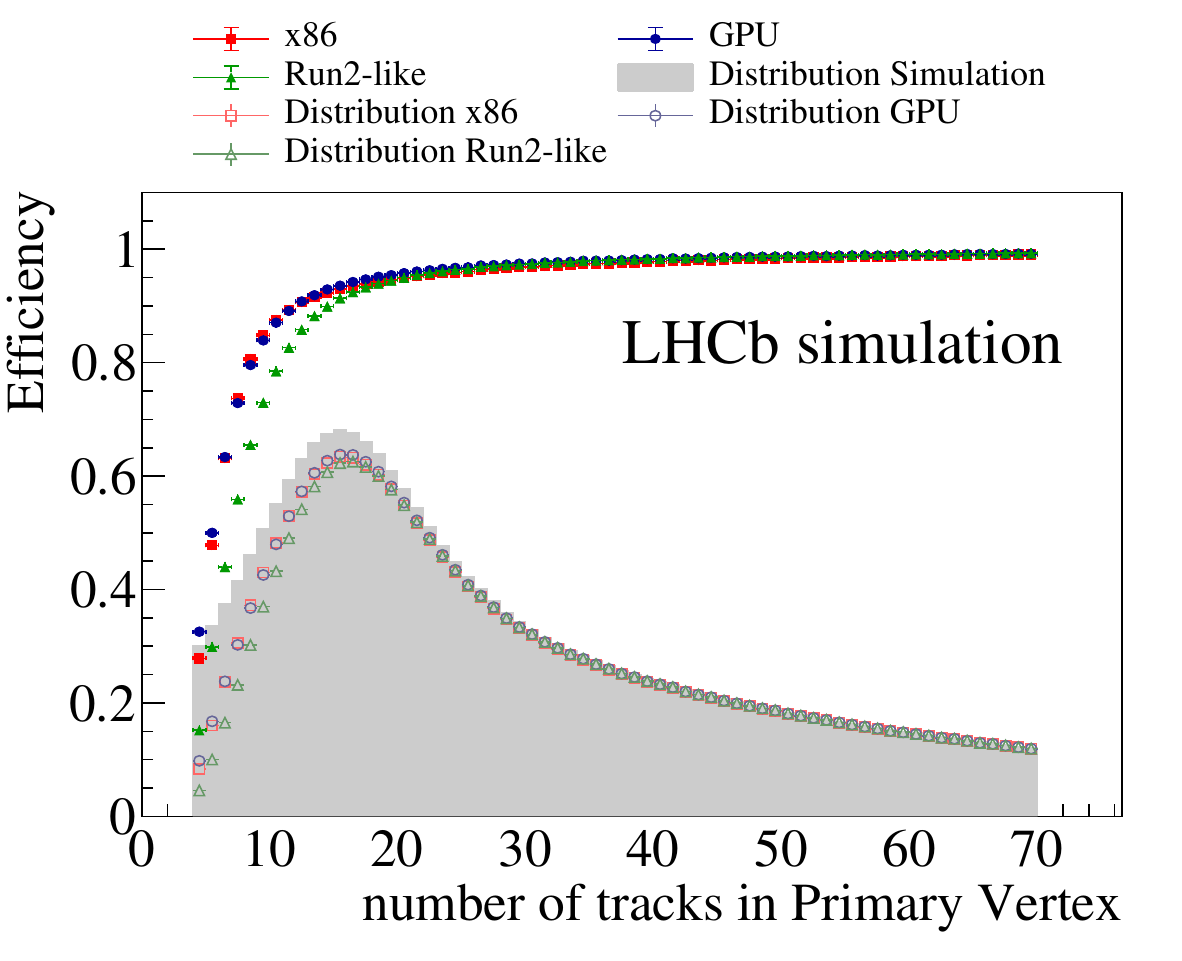}
	\includegraphics[width=0.49\linewidth]{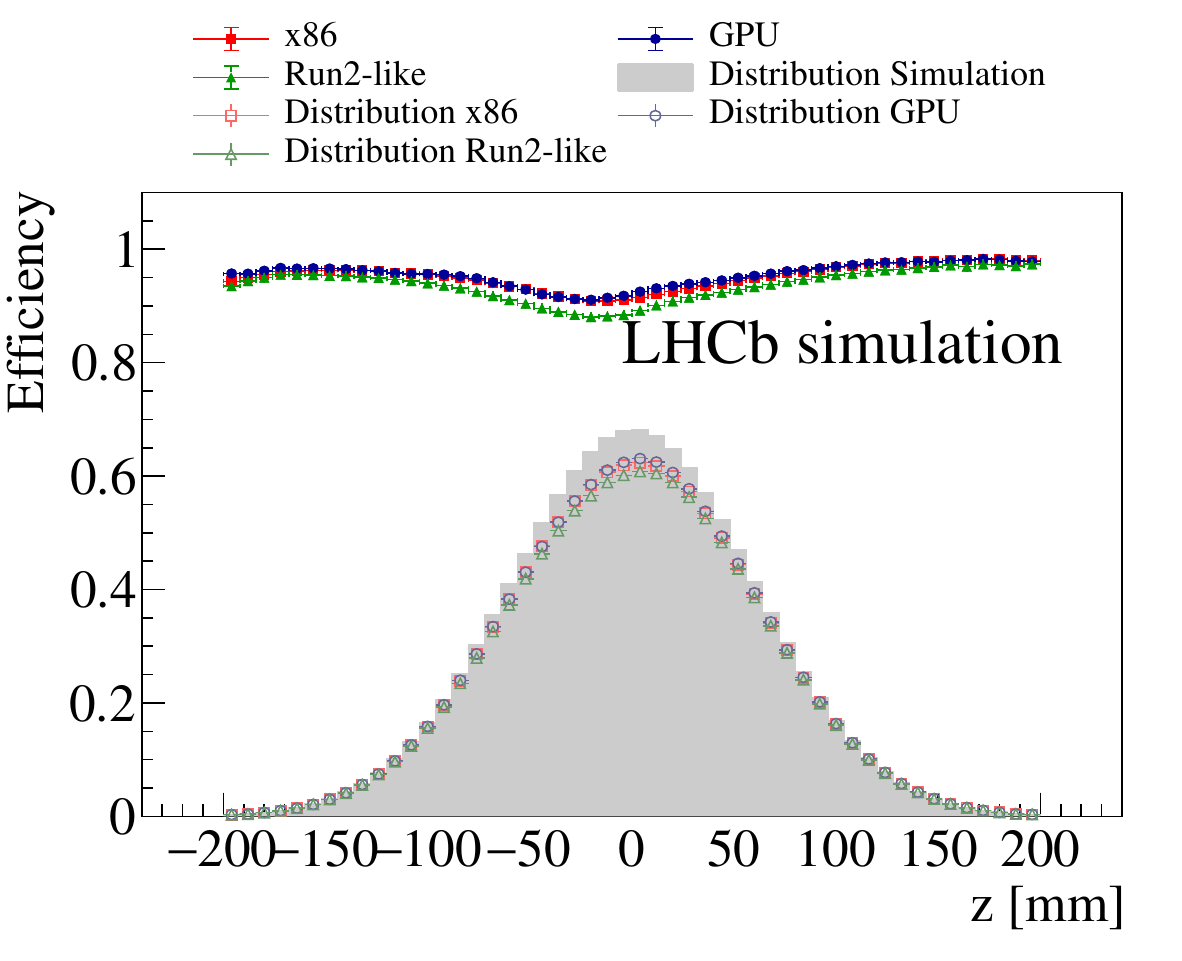}
	\caption{Primary-vertex-reconstruction efficiency as function of (left) its multiplicity and (right) its simulated $z$ position. The red squares, blue circles and green triangles points are obtained using the dedicated x86, GPU and Run2-like implementation, respectively, the grey histograms show the distribution of simulated primary vertices and the hollow red, blue, green points the number of reconstructed primary vertices in the x86, GPU and Run2-like cases, respectively}
	\label{fig:pv_comp_eff}
\end{figure*}

The PV reconstruction efficiencies for the different PV categories are summarised in Table~\ref{tab:eff}.  On average, both the x86 and GPU implementations reconstruct primary vertices with an efficiency at the level of 93.5\%, which is 2\% higher than for the Run2-like algorithm. The PV reconstruction efficiency is shown in Fig.~\ref{fig:pv_comp_eff} as a function of the number of tracks and  $z$ position in the simulated $pp$ collisions. The efficiency is expected to be lower for PVs with a smaller number of associated tracks, since the peak resulting from such PVs may not be significant enough to be identified by the peak-finding procedure. This is confirmed by the numbers in Table~\ref{tab:eff}: the PV with the highest multiplicity in the event is found in 99.5\% of the cases, while the 5th PV in multiplicity order is only found in about 91\% of the cases. On average, the PV efficiency is about 96\% for PVs with at least 10 associated tracks. The PV reconstruction efficiency is slightly reduced at the centre of the interaction region along \textit{z}. This can be explained by the observation that for those \textit{z} values PVs are more densely populated and are more likely to spatially overlap. Such PVs are harder to distinguish and could be reconstructed as a single PV instead of two distinct ones. Indeed, efficiencies of about 97\% and 89\% are found for \texttt{isolated} and \texttt{close} vertices, respectively. Both x86 and GPU implementations are highly performant for PVs which produce either beauty or charm particles (about 98\%), which are used in the majority of the physics analyses in the \lhcb collaboration. 
In comparison with Run2-like algorithm, both x86 and GPU implementations are about 17\% more efficient for finding PVs with less than 10 associated tracks and  3\% more efficient in the central $z$ region between $-50$ and $50\mm$. 

\begin{figure*}
	\centering
	\includegraphics[width=0.49\linewidth]{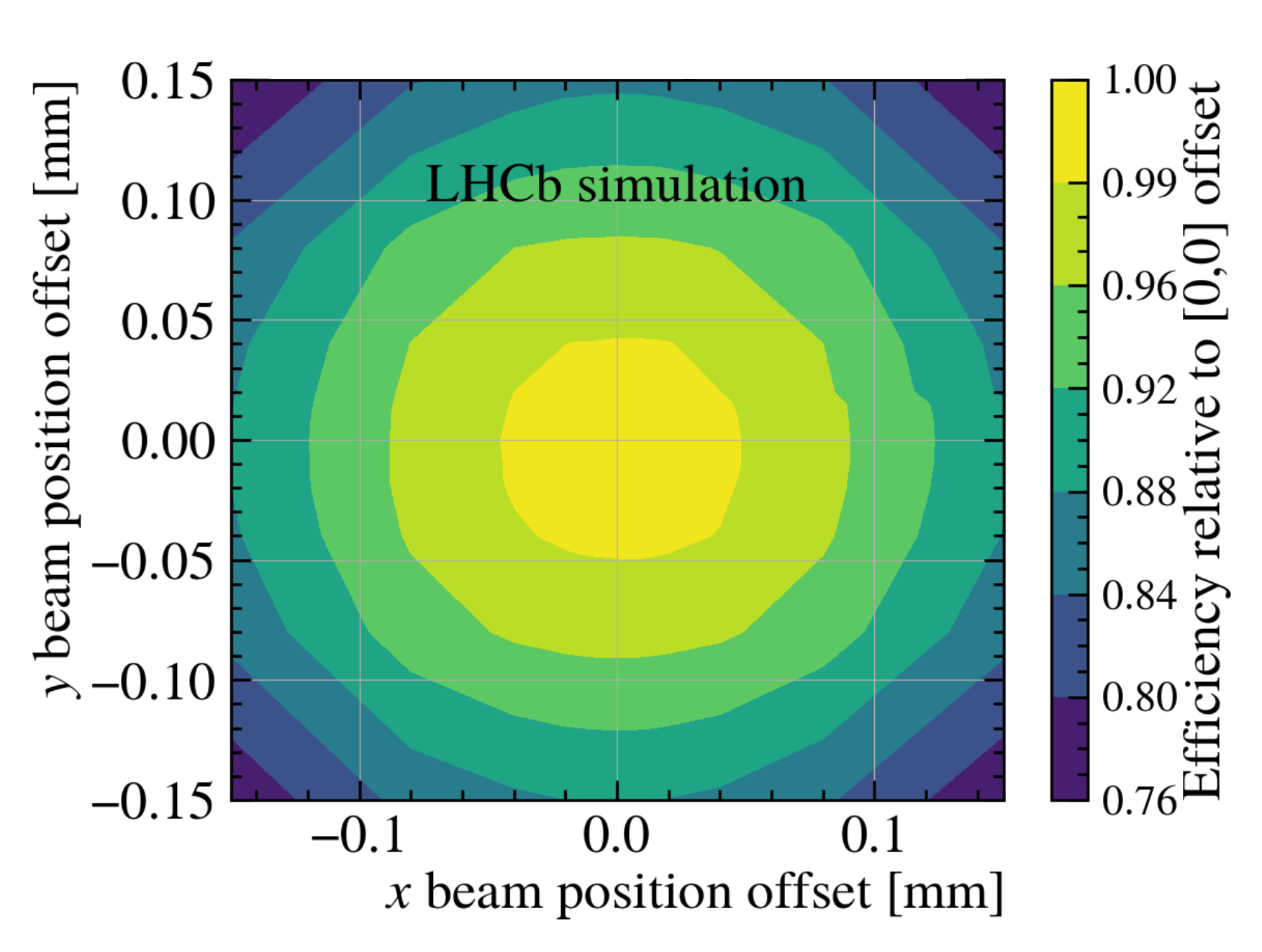}
	\includegraphics[width=0.49\linewidth]{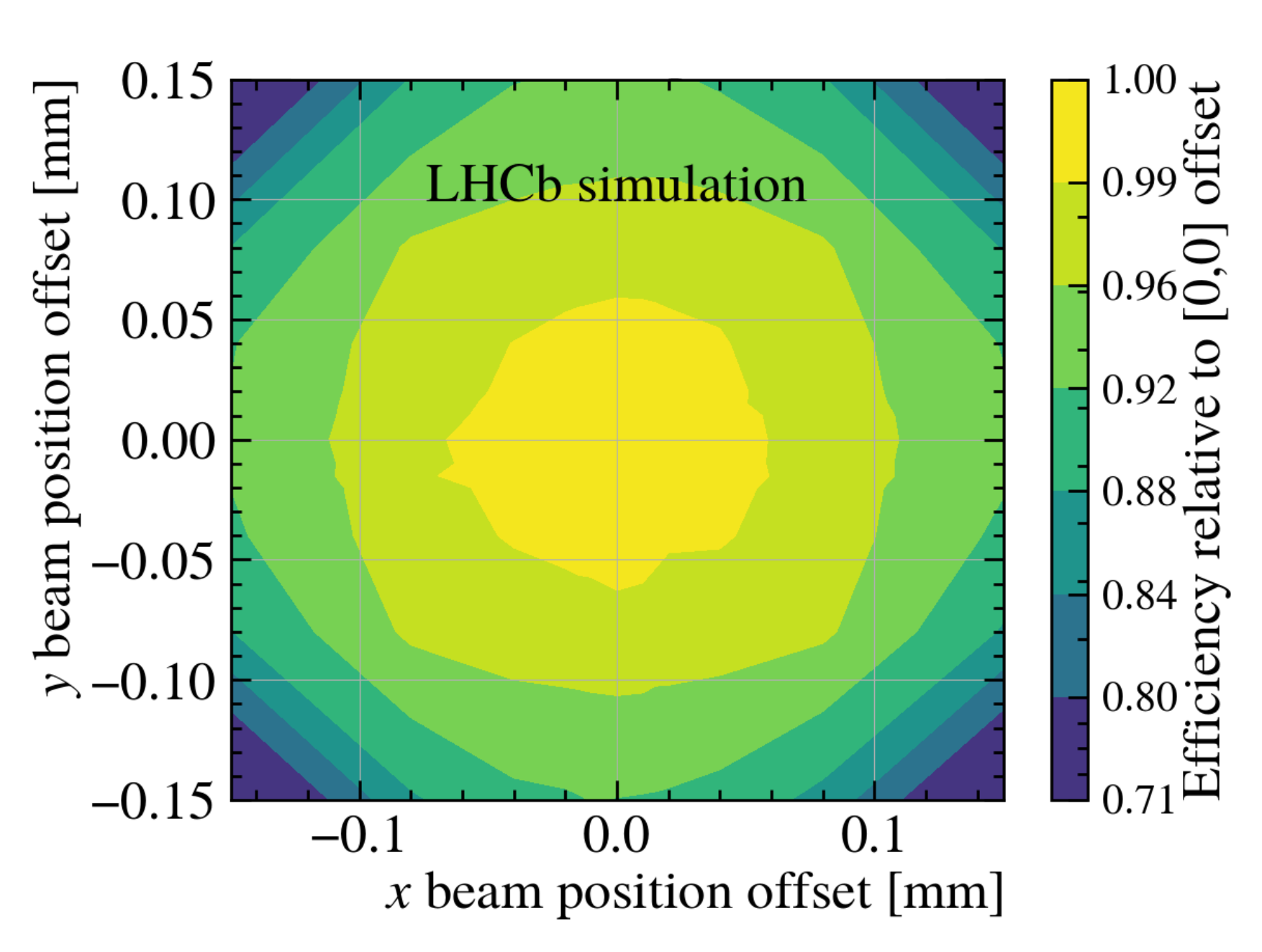}
     \caption{Relative primary-vertex-reconstruction efficiency for the (left) GPU and (right) x86 algorithm implementation as a function of beam position offsets in $x$ and $y$ directions}
	\label{fig:pv_eff_offset}
\end{figure*}

The peak-finding procedure is based on the $z$-coordinate of the point of closest approach to the beamline, making it reliant on accurate beamline position measurements. Both GPU and x86 algorithms demonstrate robustness, maintaining an efficiency of 99\% within a beamline position uncertainty of 50 $\mu$m, as illustrated in Fig.~\ref{fig:pv_eff_offset}. The efficiency remains at 99.9\% within a beamline position uncertainty of 20 $\mu$m, which is selected as the threshold for the beamline position calibration.

The measured fake rate is 1.7\% (1.6\%) for the x86 (GPU) implementation, respectively, considering all reconstructed PVs. It reduces to 0.2\% (0.6\%) for those with at least 10 associated particles. The majority of false PVs belong to the \texttt{close} category for which the fake rates are around 2.5\% for both the x86 and GPU implementations. Primary vertices with a smaller number of associated particles and which produce neither beauty nor charm hadrons are more likely to be misidentified. The comparison with Run2-like algorithm shows similar fake rate pattern. 

\begin{figure*}
	\centering
	\includegraphics[width=0.49\linewidth]{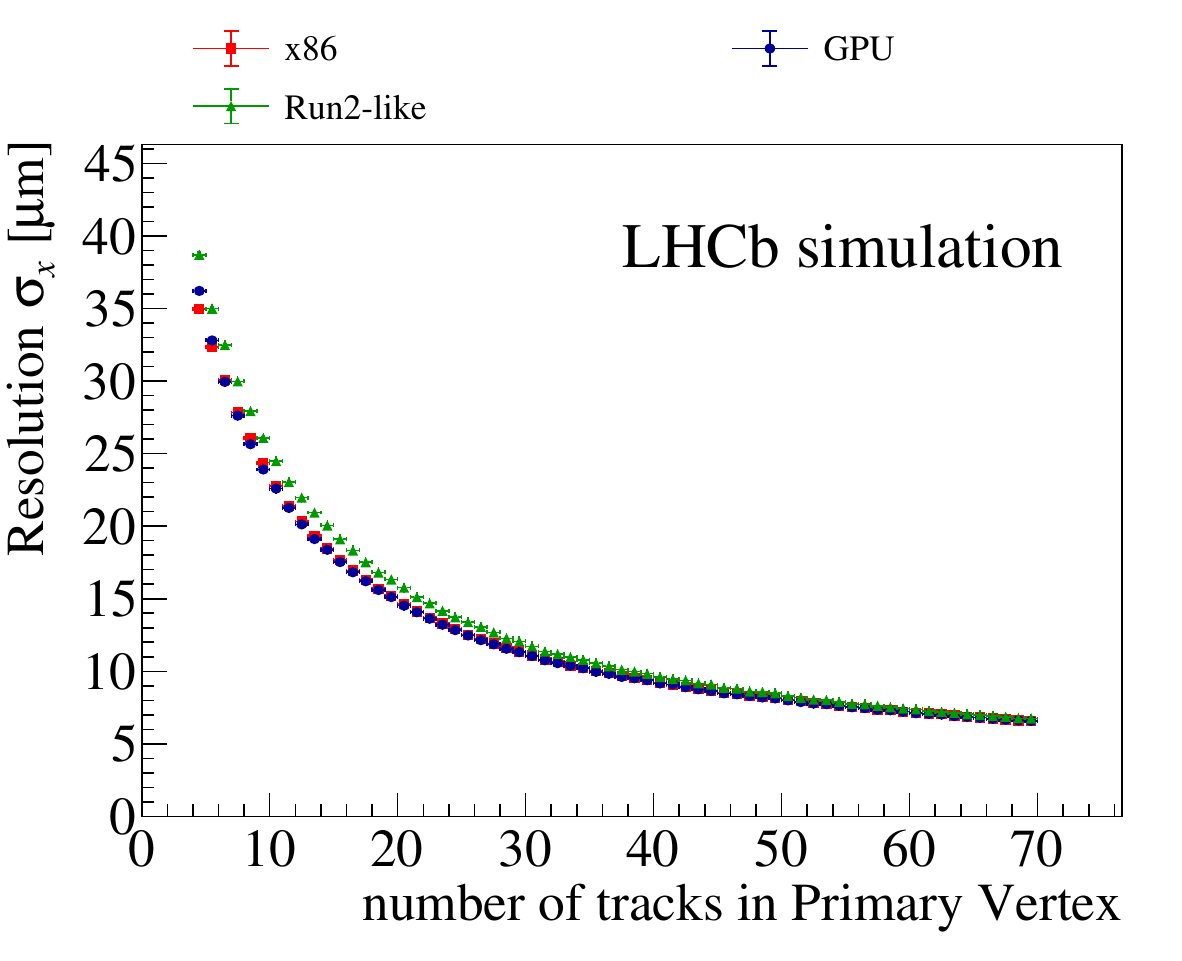}
	\includegraphics[width=0.49\linewidth]{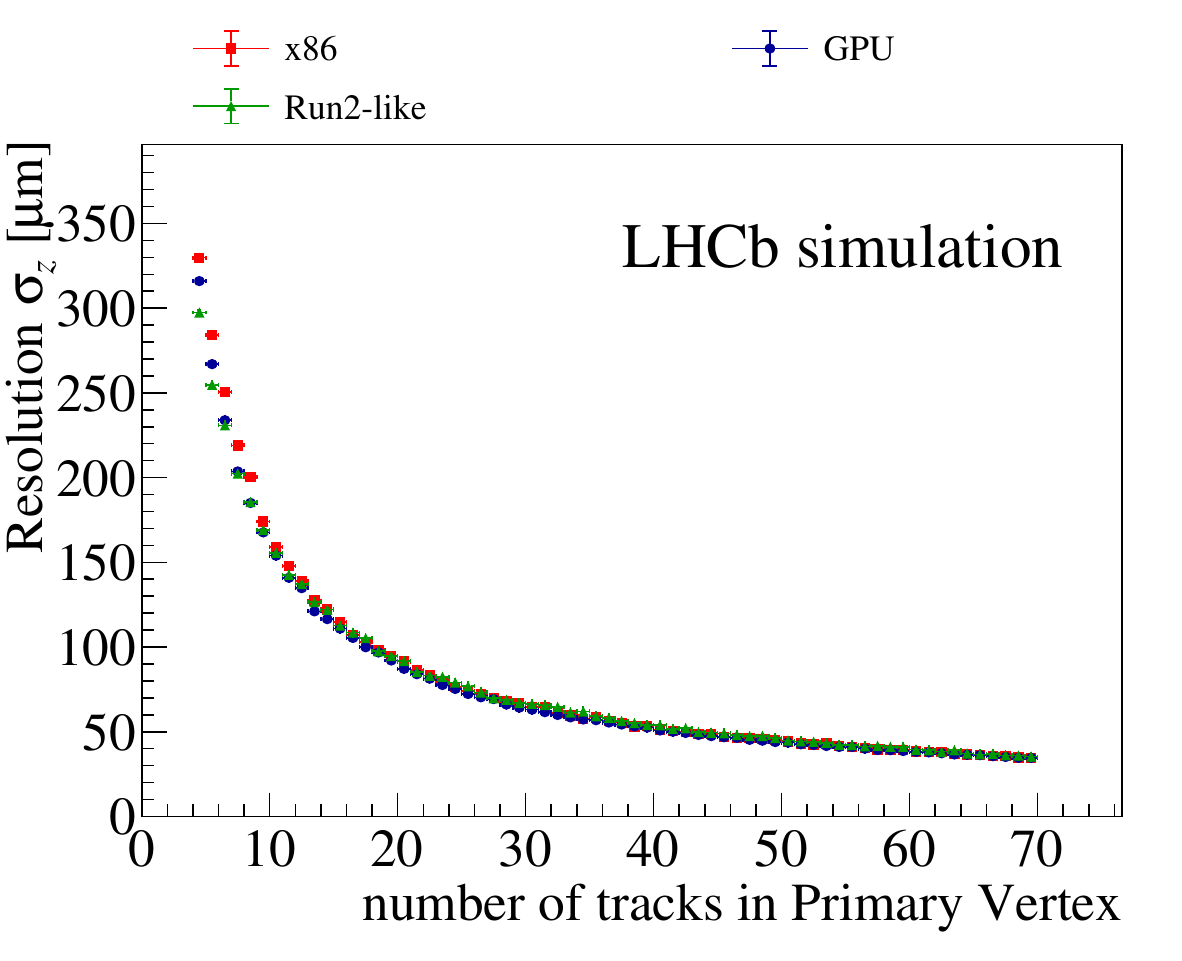}
	\includegraphics[width=0.49\linewidth]{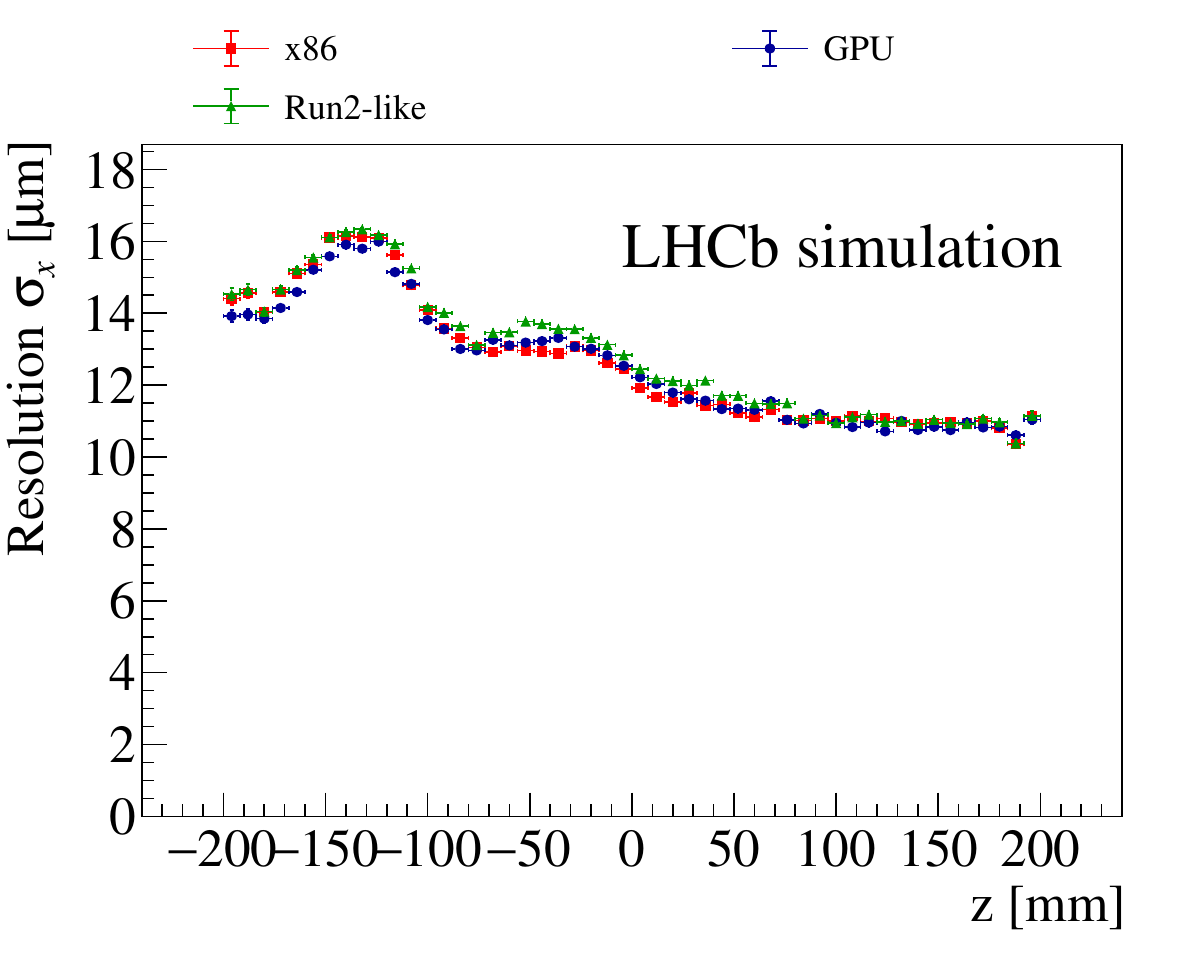}
	\includegraphics[width=0.49\linewidth]{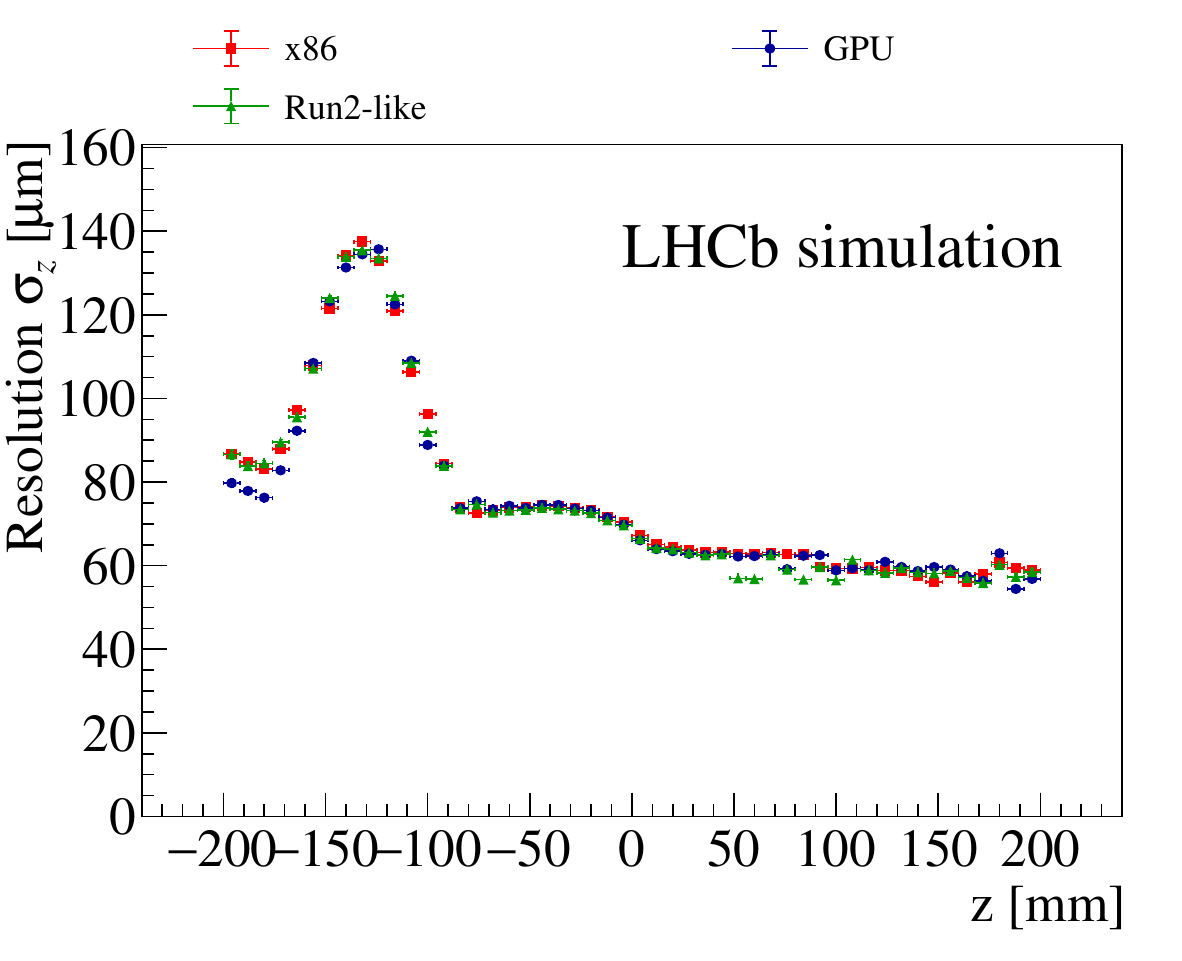}
	\caption{ Primary-vertex resolution for (left) \textit{x} and (right) \textit{z}-coordinate as (top) function its multiplicity and (bottom) the simulated primary-vertex \textit{z} position. The red squares, blue circles and green triangles points are obtained from the x86, GPU and Run2-like implementation of the primary-vertex-reconstruction algorithm, respectively}
	\label{fig:pv_comp_res}
\end{figure*}

The PV resolutions as a function of the number of tracks in the associated simulated PV and the $z$ position are shown in Fig.~\ref{fig:pv_comp_res}.
The resolution strongly depends on the number of associated particles and degrades for $-150<z<-100\mm$. This is a consequence of the \velo module spacing shown in Fig.~\ref{fig:pvinteractionregion}. The region  $-150<z<-100\mm$ falls within a gap between the detector layers, where the distance to the closest measured point is larger compared to other regions in $z$. 
While this degradation in resolution is quite significant, it should be noted that  only a small fraction of  the total number of PVs are produced in this $z$-region so the overall effect on the selection of displaced tracks is small.
The x86 and GPU implementations show a similar resolution as the Run2-like algorithm for the \textit{z}-coordinate and an improvement of $5\%$ for the \textit{x}- and \textit{y}-coordinates.

The pull distributions for all implementations show that the PV estimator is unbiased, and uncertainties are well estimated. No dependence is found for the pull mean on either the number of associated tracks in the PV or the $z$ position. Both the x86 and GPU implementations exhibit a similar dependence of the pull width on the number of associated tracks, following a pattern comparable to the resolution shown in Fig.~\ref{fig:pv_comp_res}. No dependence is found for the $z$ position. 

The reconstructed PVs in the x86 and GPU implementations are also mutually tested, considering matched PVs as those reconstructed with a smaller distance than three times their combined uncertainty. About 98\% of PVs are matched positively, with a correlation between matched PVs $x$ and $z$-coordinates of 94.5\% and 99.9\%, respectively. For the $x$-coordinate, the correlation increases to 97\% for PVs with at least 10 associated tracks.

A fraction of \velo tracks falls within the acceptance of the rest of the \lhcb detector, allowing their momentum to be determined with a precision of $0.5\text{-}1.0$\% for momenta in $2\text{-}100$\gevc. These tracks are the primary inputs to \lhcb physics analyses, and since their momentum is known, their covariance matrices are more accurate than those of the other \velo tracks. For this subset of tracks, the impact of using the more accurate track parameters and covariance matrices on the PV reconstruction performance has been evaluated with the dedicated x86 implementation. A relative improvement up to 3-5\% is seen for the low multiplicity PV resolution in $x$-direction. A relative improvement below 1\% is obtained for $z$. In the range of  $1/\pt<1$ $c$/\gev a difference up to 3-5\% is observed, but the track IP $\chi^2$ is found to agree very well with the baseline approach. The PV reconstruction efficiencies are not affected. The overall impact of this choice is therefore found to be negligible for the vast majority of use-cases, and the simpler baseline approach of treating all tracks equally is retained.

\subsection{Fixed-target primary vertex reconstruction}
In view of the simultaneous acquisition of $pp$ and $p$-gas collisions,  the PV reconstruction performance is also studied on events including collisions between \lhc beam protons and nuclei at rest in the SMOG2 target. The topology of these collisions differs to a large extent from the $pp$ case, as they occur upstream of the baseline interaction region. The lower centre-of-mass energy of $p$-gas collisions produces PVs that have a lower average track multiplicity and the created  particles are boosted in the forward direction because of the asymmetric momentum in the laboratory frame between the beam and the target. This results in a larger uncertainty when extrapolating the \velo tracks towards the beamline. The PV resolution is thus expected to be significantly worse.

\begin{figure*}
\centering
\includegraphics[width=0.48\textwidth]{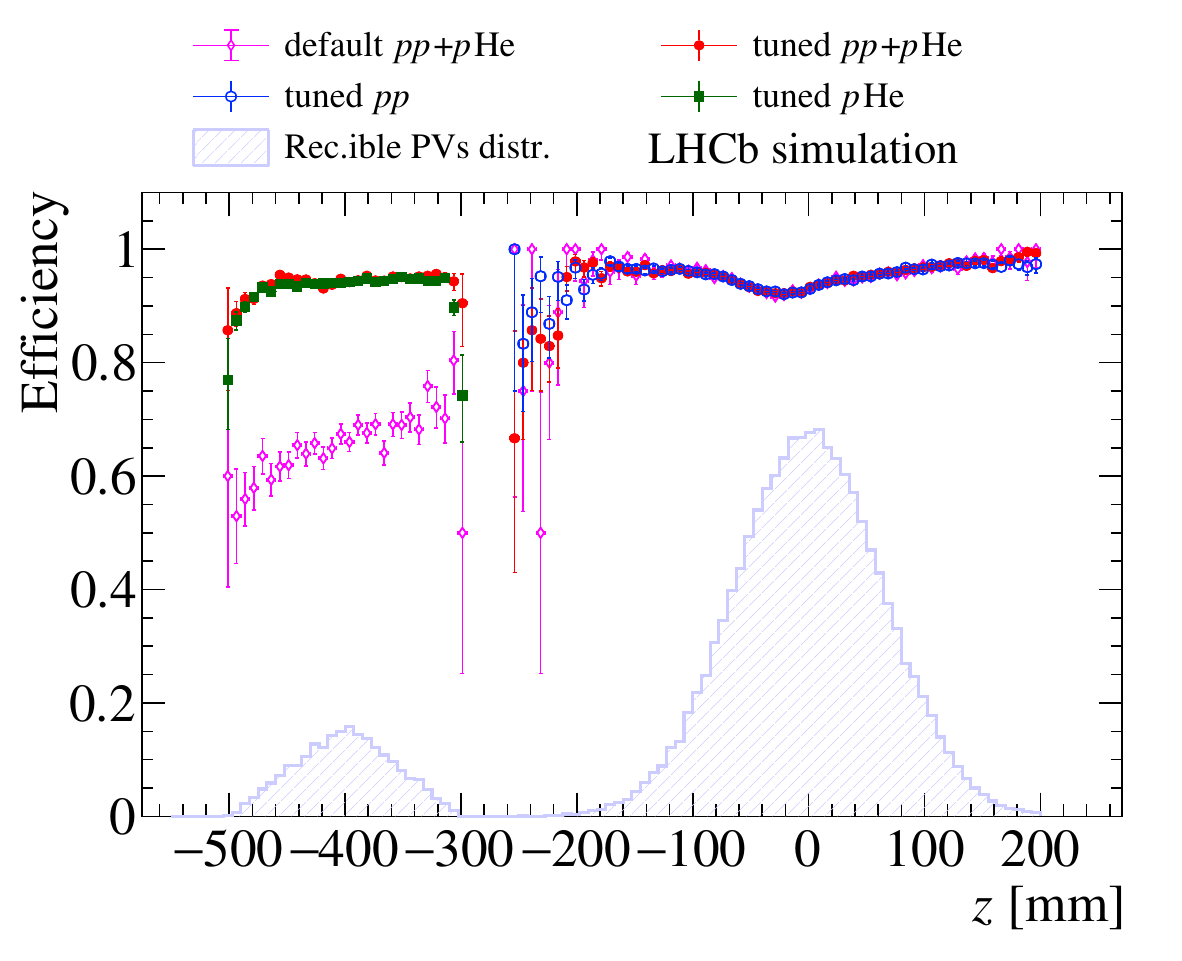}
\includegraphics[width=0.48\textwidth]{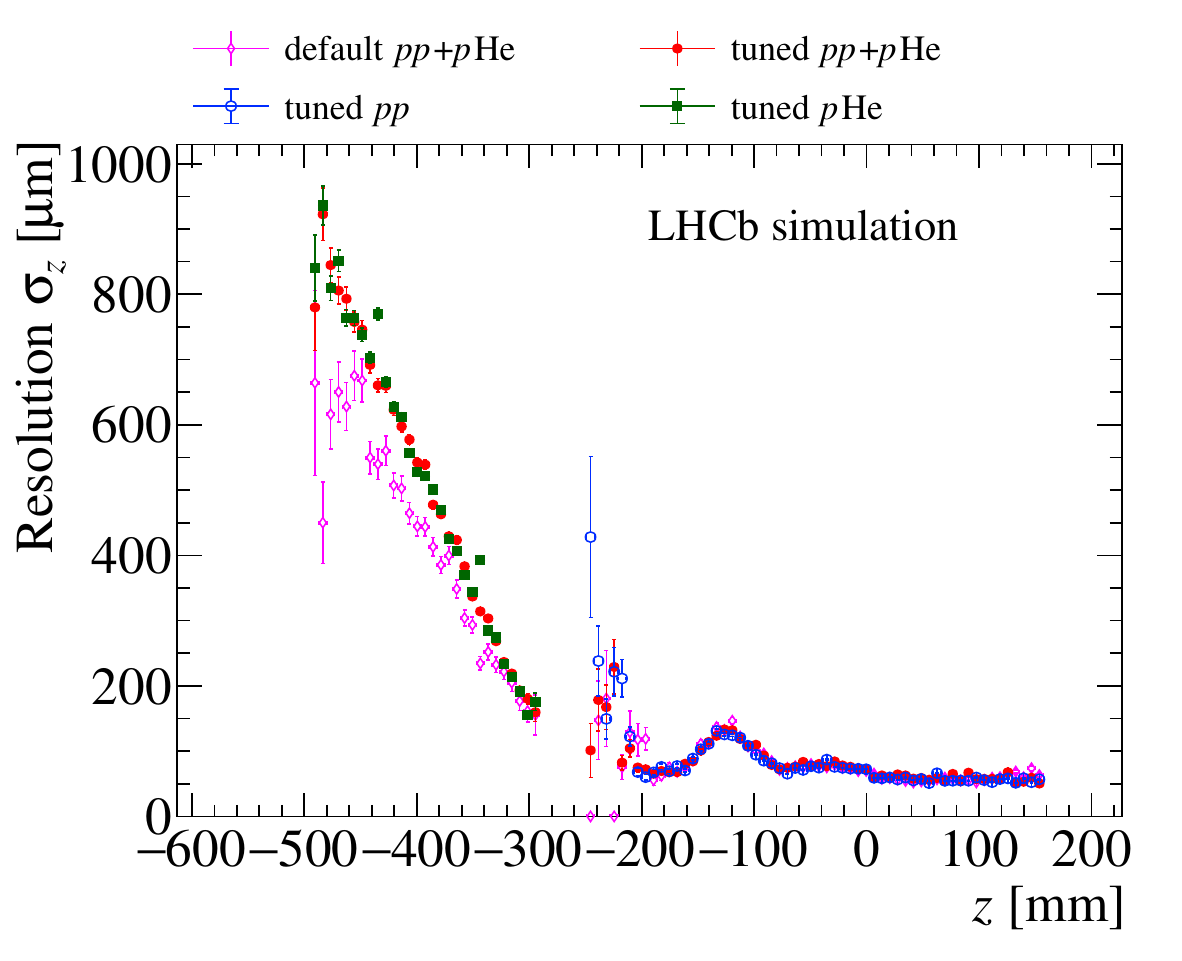}
\caption{Primary-vertex-reconstruction (left) efficiency and  (right) \textit{z} resolution  as a function of the \textit{z} coordinate of the simulated primary vertex. In both plots, the magenta curve refers to the $pp+p$He sample with the \textit{pp}-optimized algorithm implementation, while the tuned performance is shown in green, blue and red for the $pp$, $p$He and $pp+p$He samples, respectively. The longitudinal profile of the simulated positions for reconstructible vertices is also shown, on arbitrary scale, in the left plot}
\label{fig:pp_SMOG2_comp}
\end{figure*}

The algorithm reconstruction efficiency and resolution are studied on  simulated samples with three different conditions:
\begin{itemize}
	\item[$\bullet$]stand-alone \textit{p}He collisions in the SMOG2 cell; 
	\item[$\bullet$]stand-alone \textit{pp} collisions in the nominal Run~3 conditions;
	\item[$\bullet$]overlapped \textit{pp} and \proton-gas collisions with injected helium or argon as examples of light or heavy target gases. As the rate of $p$-gas collisions in the simultaneous data-taking scenario is not expected to exceed 0.2 per beam crossing, events are simulated with a single $p$-gas interaction. 
\end{itemize}
In the simulation, the gas is assumed to be confined in the region $z \in [-500, -300] \mm$~\footnote{The considered simulations assume the design position of the SMOG2 cell, differing about 4\cm from the installed one. The conclusions discussed in the following remain valid.}, with a triangular longitudinal profile (see Fig.~\ref{fig:pp_SMOG2_comp}), according to a simplified model of the expected pressure profile within the SMOG2 storage cell.
By comparing the performance in the three samples, the effect of the presence of the $p$-gas collisions on the \textit{pp} reconstruction performance, and vice versa, is assessed. 

\begin{table}
\caption{Optimisation of the primary vertex reconstruction for $pp$ and $p$-gas collisions.}
\label{tab:smog-pp}
\begin{tabular}{lll}
\hline\noalign{\smallskip}
Parameter & $pp$ & $p$-gas \\ 
\noalign{\smallskip}\hline
\textit{z} [mm]   & [-300, 300] & [-500,-300] \\
Min. tracks in the PV & 4 & 3 \\ 
Min. cluster integral  & 2.5 & 1.75 \\
Max. track $\sigmaz$  & 1.5 & 10 \\
\noalign{\smallskip}\hline
\end{tabular}
\end{table}

The PV reconstruction efficiency and resolution with the \textit{pp}-optimized implementation of the algorithm on simulated overlapped $pp$ and $p$He are shown as the magenta curves of Fig.~\ref{fig:pp_SMOG2_comp} as a function of the PV  \textit{z} position. When running the algorithm with optimal $pp$ settings, the efficiency for $p$-gas vertices is significantly lower and steeply decreases with \textit{z}. The reason are the tight thresholds set in the histogramming and clustering phases of the algorithm, optimising the speed and the physics performance for \textit{pp} collisions. A different tuning, summarised in Table~\ref{tab:smog-pp}~\cite{CERN-THESIS-2021-313}, is hence defined for $p$-gas and applied to the only vertex candidates with $z<\text{-}300 \mm$, in order to not affect the PV reconstruction performance for \textit{pp} collisions. As shown in  Fig.~\ref{fig:pp_SMOG2_comp} in red for overlapped $pp$ and $p$He simulated collisions, the specific tuning removes the inefficiency, and the algorithm is verified to provide a comparable efficiency for both types of collisions.  

\begin{figure*}
\centering
\includegraphics[width=0.48\textwidth]{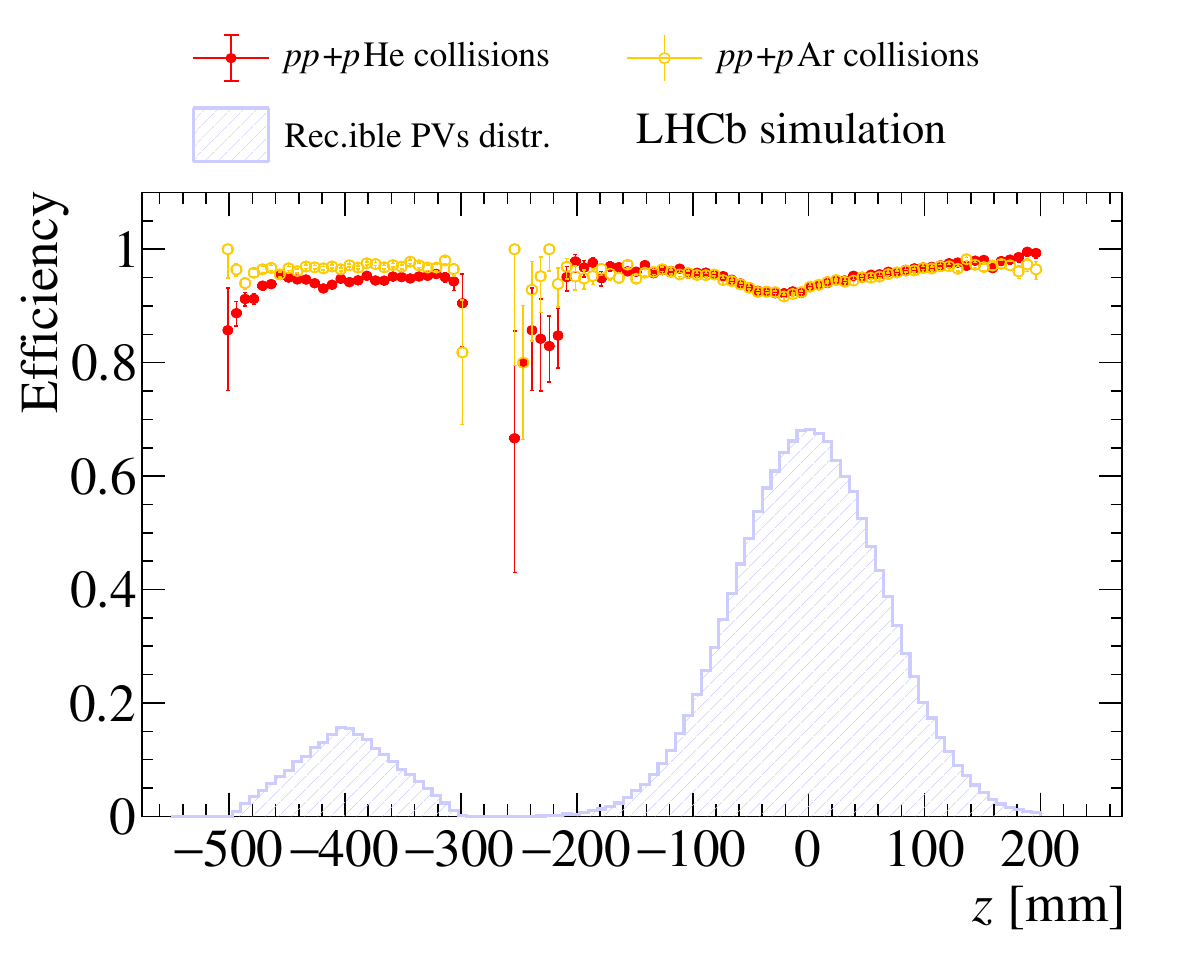}
\includegraphics[width=0.48\textwidth]{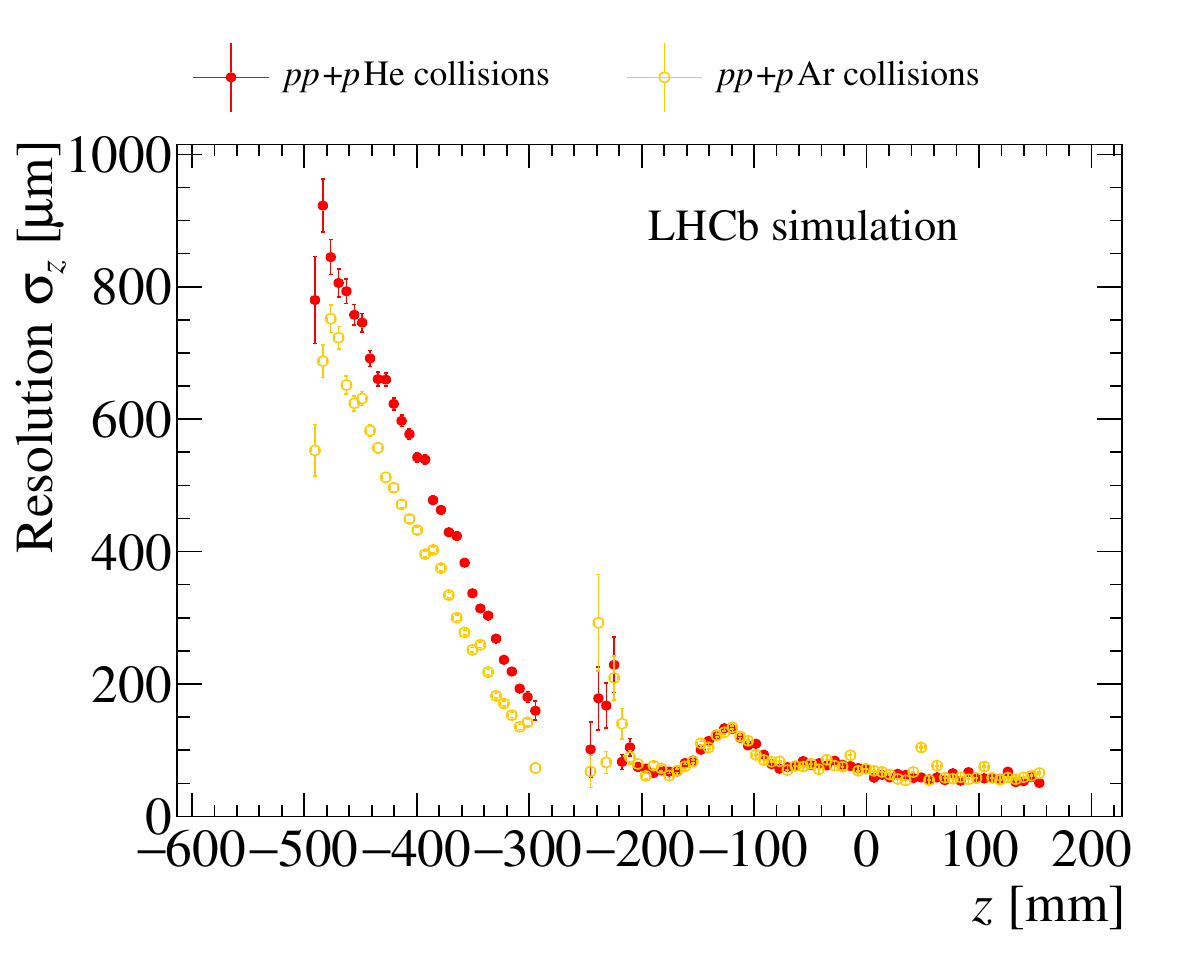}
\caption{Primary-vertex-reconstruction efficiency (left) and \textit{z} resolution (right)  as a function of the simulated PV$_z$ for the (red) $pp+p$He and (orange) $pp+p$Ar samples after the threshold tuning}
\label{fig:pp_SMOG2_HeAr}
\end{figure*}

This is not the case for the PV resolution, which steeply worsens when moving away from the central \velo region. This is expected as an intrinsic limitation due to the displaced vertex position and large track pseudorapidities in fixed-target collisions. The same conclusions are drawn when considering a heavier gas target, as shown by the performance comparison between the samples with helium or argon gas in Fig.~\ref{fig:pp_SMOG2_HeAr}, though a better performance can be seen in the argon case, as expected from the higher track multiplicity in such $p$-gas collisions.

The performance on \textit{pp} collisions is equivalent in all three conditions. This demonstrates the robustness of the reconstruction algorithm against the additional detector hits introduced by the $p$-gas collisions. The performance on $p$-gas collisions is also not affected by the simultaneous presence of \textit{pp} collisions.
Therefore, the results demonstrate that a single vertex reconstruction algorithm, configured differently for the two \textit{z} regions, achieves optimal performance for both the $p$-gas and \textit{pp} physics programs simultaneously.

\section{Vectorisation and parallelism}
\label{sec:implementation}

\subsection*{x86}
The described algorithm has sections where an operation is performed for multiple or all tracks, which is a prime candidate for vectorisation. The track class model~\cite{Hennequin:2019itm} implements a structure of arrays design, where the same data members (e.g. $x_{trk}$) of all tracks are contiguous in memory. Chunks of data members can therefore be loaded into size $N \in \{4, 8, 16\}$ vector registers quickly. With the help of vector operations~\cite{intel_intrinsics}, N tracks can then be processed at the same time. The track preparation and extrapolation steps are executed this way and matrix operations are particularly accelerated by using vectorisation. Filling of the histogram, the peak search, and the partitioning all cost little time and are executed sequentially, since the operations there have too many interdependencies to be vectorised efficiently. 
Tracks are sorted by their partition with a vector gather operation and the vertex fit parallelises over all tracks in the same partition. 

\subsection*{GPU}
\label{sec:impl_gpu}
The GPU implementation achieves the necessary throughput by making use of the thread- and block-level parallelism. Since events are independent from each other, batches of several hundred events are processed in parallel, with one block per event for every step of the PV reconstruction. Thread-level parallelism is defined individually for every step as described in the following. 
The first step, the track extrapolation, can be performed in parallel by assigning one thread to a track since the track states are independent from each other and are read from and saved to distinct places in memory. The histogram can be filled in a similar manner, where one thread is assigned to an extrapolated track, looks up its \zpoca--position and increases the corresponding histogram bin, again taking into account the uncertainty on \zpoca. To avoid race conditions, where two threads access and write to the same memory location at the same time, atomic functions are used.

The peak search follows the same sequential logic of the initial CPU implementation. A possible optimisation would be to subdivide the histogram into different, possibly overlapping regions, where within every region a thread is assigned to identify peaks. 

The next two steps, the association of tracks to PV candidates and the PV fitting, are done for every vertex fit in parallel by one thread, and the $\chi^2$-sum over all tracks and derived quantities is again parallelised for every vertex candidate. 
To speed up the calculations and to prevent completely unrelated tracks, whose weights would be almost zero, from contributing to the vertex fit, only tracks within a certain $\chi^2_i$ with respect to the vertex candidate are considered. To speed up the matrix calculations, they are explicitly written out exploiting the fact that many elements are zero.

\section{Throughput performance}
\label{sec:throughput}

\begin{figure*}
\centering
\includegraphics[width=0.75\textwidth]{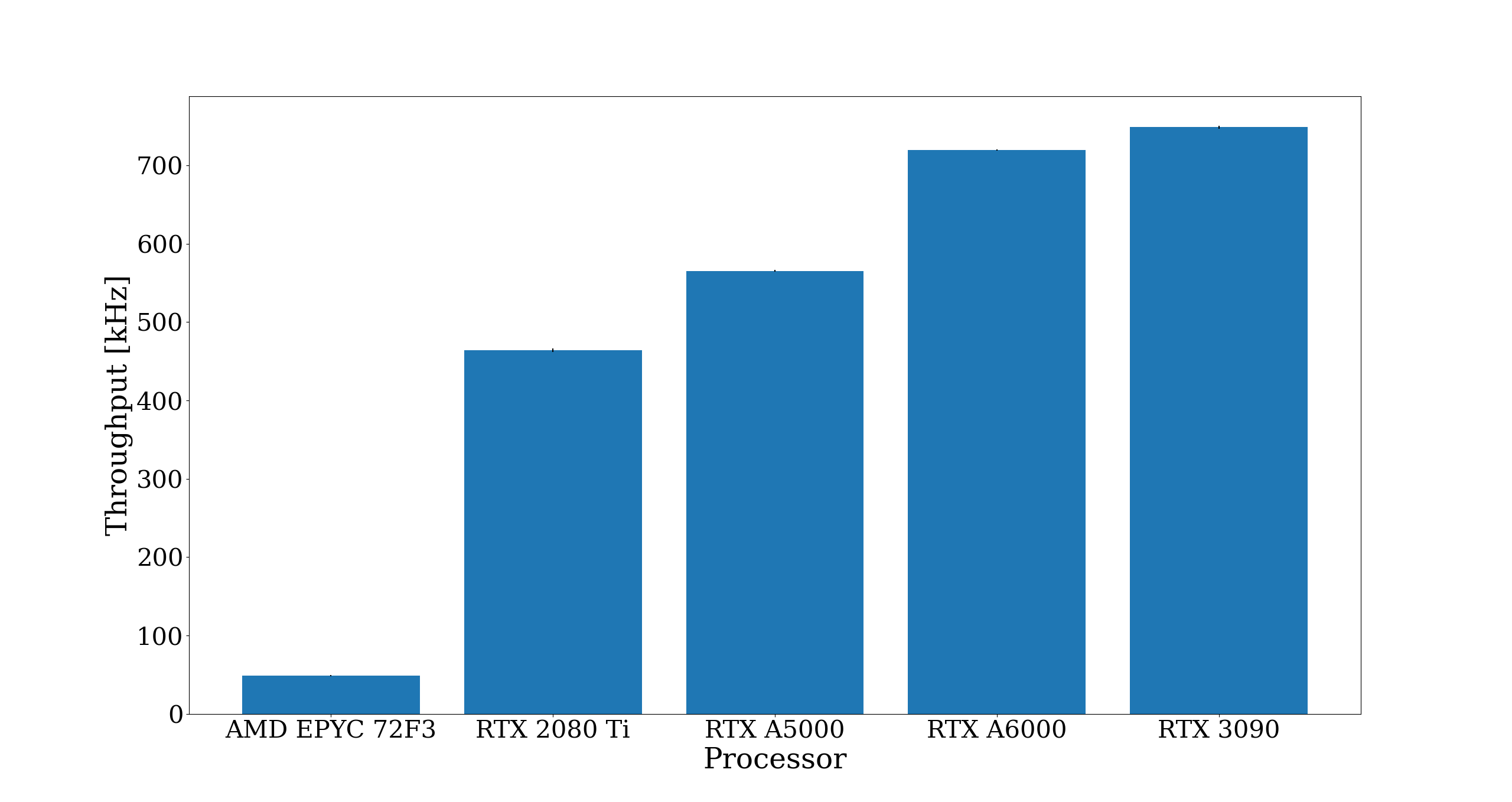}
\caption{Throughput of the algorithms optimised for GPU architecture on various GPU cards and of the x86 one on an AMD EPYC 72F3 server. This includes the preprocessing algorithms producing input to the primary vertex finding and the primary vertex finding algorithm itself. The relative measurement uncertainty of around 0.2\% is too small to be seen in the figure}
\label{fig:pv_throughput}
\end{figure*}

\begin{figure*}
\centering
\includegraphics[width=0.46\textwidth]{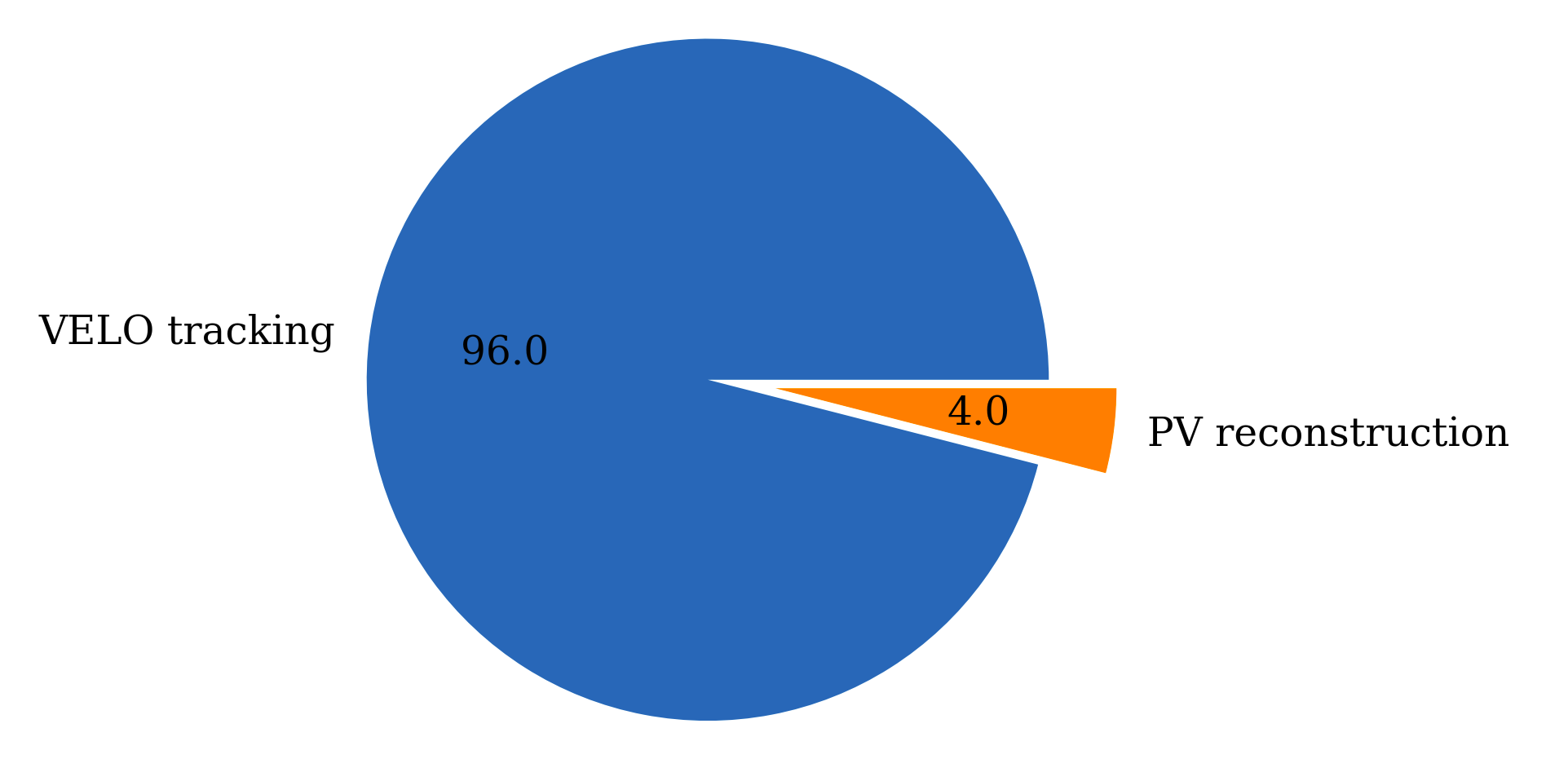}
\includegraphics[width=0.50\textwidth]{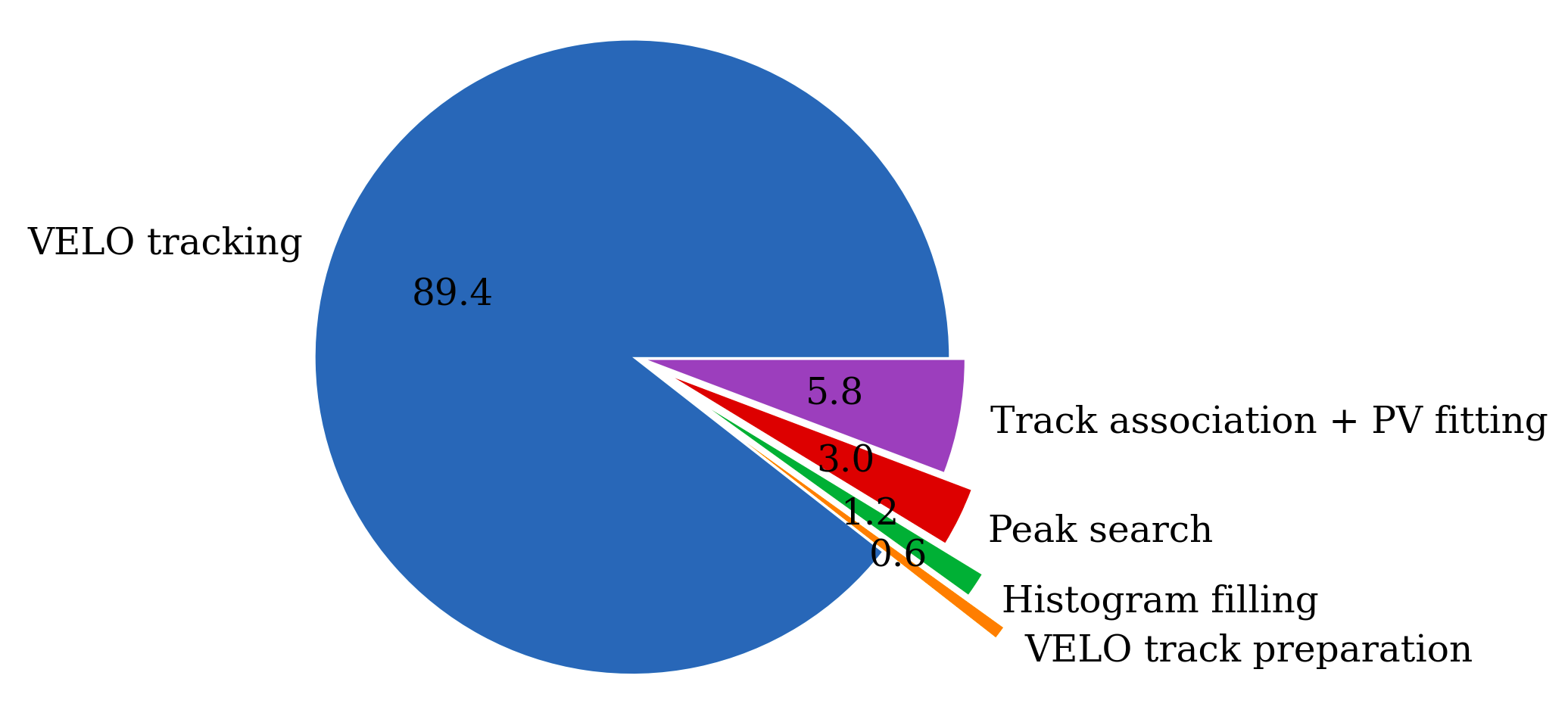}
\caption{Breakdown of the primary vertex reconstruction sequence optimised for (left) x86  and for  (right) a RTX A5000 GPU architecture. The primary vertex finding algorithms adds up to 4\% and 10\% of the total processing time, respectively. For the GPU architecture, the time spent in every step of the algorithm is individually measured and the primary vertex fit dominates}
\label{fig:pv_algo_breakdown}
\end{figure*}

The throughput of the PV reconstruction sequence is measured on a CPU and several GPU cards. This includes the \velo raw data decoding, clustering of individual pixel measurements, \velo tracking and the PV reconstruction itself. It should be noted that the pre-processing algorithms have different implementations optimised for the CPU~\cite{Hennequin:2019itm} and the GPU~\cite{VELO_GPU} architectures. Therefore, the fraction of time spent on PV finding cannot be compared directly, but gives an indication of the optimisation of the algorithm with respect to the other algorithms in the sequence. The measurement of the GPU throughput includes data transfers between the CPU and the GPU. Several CUDA streams are launched in parallel, each processing separate batches of events, to keep the GPU busy with compute operations while data transfers occur.
Figure~\ref{fig:pv_throughput} shows the throughput on the different types of hardware for simulated \textit{pp}-collision events, while Fig.~\ref{fig:pv_algo_breakdown} shows the fraction spent on the PV finding for the algorithm optimised for GPU and x86 architectures, respectively. As discussed in Ref.~\cite{LHCb:2021kxm}, both implementations, as well as all \hltone reconstruction algorithms, meet the requirement of processing 30 MHz of input data with the available resources. When processing simulated \textit{p}He-collisions, as used in \lhcb fixed-target program, the throughput on a single GPU card decreases by 5\%.

\section{Conclusion}
\label{sec:conclusion}
A new vertex finding algorithm is developed for the high-level trigger of the \lhcb Upgrade detector. It is shown to deliver sufficient physics performance while having a high enough throughput. It is demonstrated that the algorithm can be parallelised at different levels and therefore efficient implementations on both x86 and GPU architectures are possible. Both the x86 and GPU implementations outperform the previous Run2-like algorithm in terms of reconstruction efficiency and resolution. It is shown that the algorithm can be further extended to the beam-gas region, which is separated from the nominal $pp$ collision region. With small adjustments of the parameters of the algorithm beam-gas PVs can be reconstructed with high efficiency despite them being more difficult to treat, without disturbing the reconstruction of $pp$ PVs. This offers to the \lhcb experiment the possibility to simultaneously take beam-beam and beam-gas collision data.

\begin{acknowledgements}
We thank the \lhcb Real-Time Analysis project for its support, for many useful discussions, and for reviewing an early draft of this manuscript. We also thank the \lhcb computing and simulation teams for producing the simulated \lhcb samples used to benchmark the performance of the algorithm presented in this paper. The development and maintenance of the \lhcb nightly testing and benchmarking infrastructure which our work relied on is a collaborative effort and we are grateful to all \lhcb colleagues who contribute to it. VVG, FR, and DvB were supported by the European Research Council under Grant Agreement number 724777 ``RECEPT'' during the period in which part of this work was carried out. DvB further acknowledges support of the European Research Council Starting grant ``ALPaCA'' 101040710. AD, MG and TW would like to  expresses their gratitude to the Ministry of Science and Higher Education in Poland, for financial support under the contract no 2022/WK/03. 
\end{acknowledgements}

\textbf{Data Availability Statement} Data will be made available on reasonable request. [Author's comment: The datasets generated and analysed during this study are available from the corresponding authors on reasonable request.]

\textbf{Code Availability Statement} The code/software used for this study is openly accessible under the licenses Apache-2.0 \url{https://gitlab.cern.ch/lhcb/Allen} and GPL Version 3 \url{https://gitlab.cern.ch/lhcb/Rec}.

\textbf{Open Access} This article is licensed under a Creative Commons Attribution 4.0 International License, which permits use, sharing, adaptation, distribution and reproduction in any medium or format, as long as you
give appropriate credit to the original author(s) and the source, provide a link to the Creative Commons licence, and indicate if changes
were made. The images or other third party material in this article
are included in the article’s Creative Commons licence, unless indicated otherwise in a credit line to the material. If material is not
included in the article’s Creative Commons licence and your intended
use is not permitted by statutory regulation or exceeds the permitted use, you will need to obtain permission directly from the copy-
right holder. To view a copy of this licence, visit \url{https://creativecommons.org/licenses/by/4.0/}.




\end{document}